\documentclass[reprint,amsmath,amssymb,aps,pra]{revtex4-1}

\usepackage{graphicx} 
\usepackage{dcolumn}  
\usepackage{bm}
\usepackage{color}

\begin{document}

\title{Pair correlation of atoms scattered from colliding Bose-Einstein quasicondensates}

\author{Pawe{\l} Zin$\,^{1,2}$, Tomasz Wasak$\,^{3}$, Denis Boiron$\,^{4}$ and Christoph I.  Westbrook$\,^4$ }

\affiliation{\mbox{$^1$National Center for Nuclear Research,\mbox{ul.~Pasteura 7}, PL-02-093 Warsaw, Poland} \\
\mbox{$^2$Faculty of Physics, University of Warsaw, \mbox{ul.~Pasteura 5}, PL-02-093 Warsaw, Poland } \\
\mbox{$^3$Max Planck Institute for the Physics of Complex Systems, N\"othnitzer Str. 38, 01187 Dresden, Germany}
\mbox{$^4$Laboratoire Charles Fabry, Institut d'Optique Graduate School, CNRS, Universit\'e Paris-Saclay 91127 Palaiseau cedex, France}\\
}

\begin{abstract}
A collision of Bose-Einstein condensates is a useful source of single nonclassically correlated pairs of atoms. Here, we consider elastic scattering of atoms from
elongated clouds taking into account an effective, finite duration of the collision due to the expansion of the condensates. Also, we include the quasicondensate nature
of the degenerate quantum gases, due to a finite temperature of the system. We evaluate the pair correlation function measured experimentally in
K. V. Kheruntsyan, {\it et. al.}  Phys. Rev. Lett. {\bf 108}, 260401 (2012). We show that the finite duration of the collision is an important factor determining the
properties of the correlations. Our analytic calculations are in agreement with the measurements. The analytical model we provide, useful for identifying physical processes that
influence the correlations, is relevant for experiments with nonclassical pairs of atoms.
\end{abstract}

\maketitle

\newcommand{\K}{{\bf k}}
\newcommand{\dk}{\Delta {\bf k}}
\newcommand{\DK}{\Delta {\bf K}}
\newcommand{\KK}{{\bf K}}

\newcommand{\x}{{\bf r}}
\newcommand{\dx}{\Delta {\bf r}}
\newcommand{\DX}{\Delta {\bf R}}
\newcommand{\X}{{\bf R}}

\newcommand{\B}[1]{\mathbf{#1}} 
\newcommand{\f}[1]{\textrm{#1}} 

\newcommand{\half}{{\frac{1}{2}}}

\newcommand{\vv}{{\bf v}}
\newcommand{\p}{{\bf p}}
 
\newcommand{\comment}[1]{\textcolor{blue}{#1}}

\section{Introduction}\label{Int}


Ultracold atoms offer a promising platform that is relevant for studies of the foundations of quantum mechanics and also for applications that rely on quantum effects.
The controlled generation of correlated pairs of atoms is an important method in this context, since such pairs play a similar role in quantum atomic physics as pairs of
photons in quantum optics. The latter were employed in demonstration of the violation of Bell's inequalities for photons~\cite{bellAspect}, the photonic Hong-Ou-Mandel
effect~\cite{oum} or the ghost-imaging~\cite{ghostphotons}. In the atomic context, generation of correlated pairs of atoms was
reported~\cite{tunableParis,viennaTB,keterle1,paryz0,paryz1}, and it was shown theoretically and experimentally that they can be employed for demonstrating, for example,
sub-Poissonian statistics of atoms~\cite{paryz2,wasakdeuar}, the violation of the atomic Cauchy-Schwarz inequality~\cite{paryz3,wasakCSI, wasakCSI2, wasakEnt},
Hong-Ou-Mandel effect for atoms~\cite{atomicHOMtheo,atomicHOM}, or atomic ghost-imaging~\cite{ghost,ghost2}. The nonclassicality of the spatially separated, correlated
atomic pairs has recently been employed for demonstration of Bell correlations~\cite{wasakBellCorr}, and for 3D magnetic gradiometry~\cite{gradioTruscott}. Furthermore,
the entangled pairs could find applications for the violation of Bell's inequality for massive particles~\cite{atomBell, wasakBell}, atomic
interferometry~\cite{njpWasak}, study of the quantum signatures of analog Hawking radiation~\cite{hawkingZin}, or fundamental experimental tests of quantum
mechanics such as Einstein-Podolsky-Rosen gedanken experiment~\cite{Wieden}.

In our work, we focus on a particular method of generating correlated pairs of atoms. In this scenario, the pairs are emitted from collisions of counter-propagating
ultracold degenerate atomic Bose gases~\cite{paryz0}.  As a result of binary collisions between the particles that constitute the counter-propagating clouds,
atomic pairs scatter out from the clouds with opposite velocities.  In the spontaneous regime, where bosonic enhancement does not influence single collision events, the
direction of velocity of outgoing particles is random.  Due to the superposition principle, the quantum state of single atomic pair is entangled in different momentum
directions~\cite{Wieden}.

To further exploit the generated pairs of atoms, a detailed control over the produced state of atomic pairs is required. To certify this control, some preliminary
measurements are helpful. A particularly important example is provided by the observation of the correlation functions between pairs of atoms. Such properties of atoms
scattered from colliding ultracold clouds was considered in the literature~\cite{jur,bach,zin1,zin2,zin3,zin4,deuar1,deuar2,gardiner1,gardiner2,paryz10,karen0,karen1,
  wasakdeuar,zinnowy}.  In particular, in the previous paper~\cite{zin4}, we described an analytic calculation of pair correlation functions.  Some of the results were in
insufficient agreement with experimental results.

Since this previous work~\cite{zin4}, we have refined and improved both the calculations and the experiment.  From the experimental point of
view, we have improved the signal to noise ratio, and simplified the collision geometry~\cite{paryz3}.  On the theory side, we are now able to take into account the
expansion of the condensate during the collision, and show that the condensate expansion reduces the atom density and, therefore, also the collision rate.  Thus, this
effect limits the duration of collision and emission of pairs, and this finite time leads to an energy broadening of the decay products.  We show that this broadening
significantly improves the agreement between  theory and experiment. Also, we take into account the fact that in reduced geometries, when the system is highly
elongated, the phase of a degenerate Bose gas fluctuates on a scale shorter than the dimension of the cloud, resulting in a quasi-Bose-Einstein condensate.
Finally, we include into the formalism the interaction of the scattered atoms with the mean-field potential of the colliding clouds, a factor that is often neglected.
With our analytical treatment, we can identify the physical processes which affect the properties of the correlation functions.


The paper is organized as follows.
In Sec.~\ref{sec1}, we introduce the experimental context, see Sec.~\ref{sec1a}, 
and the method used to describe the quasicondensate, see Sec.~\ref{sec1b}, and calculate its most important properties.
In Sec.~\ref{sec2}, we introduce the theoretical description of the quasicondensate collisions.
Here, we introduce the variational ansatz that describes the evolution of the counter-propagating quasi-condensates.
In Sec.~\ref{sec3}, we present the description of the scattered atoms based on the Bogoliubov method.
There, we give analytical formulas for the pair correlation function.
The pair correlation function contains two types of correlations, which we shall label as ``local'' and ``opposite''.
The local correlation involves atoms with nearly parallel momenta and is directly related to the single particle correlation function.
The opposite correlation involves the creation of pairs with nearly antiparallel momenta. 
In the previous papers~\cite{paryz2,paryz3}, we used the labels ``collinear'' and ``back to back'' instead of ``local'' and ``opposite'', respectively.
In Sec.~\ref{sec4}, performing a series of controlled approximations, and using a Gaussian variational ansatz, we derive formulas for 
the single particle correlation function and local part of the pair correlation function.
In Sec.~\ref{sec5}, proceeding in a similar way as in Sec.~\ref{sec4}, we derive formulas for the 
opposite part of the pair correlation function.
In Sec.~\ref{sec6}, we apply the formulas obtained in the previous sections,
and provide theoretical calculations of the normalized pair correlation function that is measured in the experiment.
We compare the theoretical and experimental results.
We close this paper in Sec.~\ref{sec7} with a summary.
The technical calculations are moved to appendices.


\section{Quasicondensate description}\label{sec1}

\subsection{Experimental setup} \label{sec1a}

In the experiment~\cite{paryz3}, we had approximately  $N\approx10^5$ metastable helium atoms at temperature 
$T_s \approx 200 \, \mbox{nK}$
placed in a strongly elongated harmonic trapping potential
\begin{equation}
\label{pot}
V(\x) = \frac{1}{2}m\big[\omega_r^2(x^2+y^2) + \omega_z^2 z^2\big],
\end{equation}
where $\omega_r=1500 \times 2\pi\, \mbox{Hz}$
and $\omega_z= 7.5 \times 2\pi \, \mbox{Hz} $. 

After trap switch-off, the cloud is transferred into $m_F =0$ hyperfine state, and divided equally into momentum wave packets centered at velocities $v_0=
9.2 \, \mbox{cm/s}$ and $-v_0$.  As a result of binary collisions within the counter-propagating cloud, the atoms are scattered out and form a halo in momentum space.
After $46$~cm of free fall, the atoms fall onto a micro-channel plate (MCP) detector, which records  two-dimensional positions and  arrival times of individual atoms.
This information allows us to determine the pair correlation function of the cloud of scattered particles.  The precision of the measurement is limited by a finite
resolution of the MCP.  The resolution will be taken into account in our comparison between the theoretical estimates and the experimental results.

\subsection{Bogoliubov method} \label{sec1b}

Since the system is elongated along the $z$-axis, we need to include the description of the phase fluctuations of the BEC.  To this end, we divide the field operator into
two parts $\hat \Psi = \hat \Psi_{qc} + \hat \delta $ where $\hat \Psi_{qc}$ describes the quasicondensate and $\hat \delta $ the scattered atoms.  We describe the
quasicondensate within Bogoliubov method in the density-phase representation \cite{gora}, where
\begin{equation}
  \label{psi_qc}
  \hat \Psi_{qc} = e^{i \hat \phi} \sqrt{\hat n}  = e^{i \hat \phi} \sqrt{ n + \delta \hat n }
\end{equation}
where $n(\x) = \langle \hat n(\x) \rangle$ is the mean density given by the solution
of the Gross-Pitaevskii (GP) equation
\begin{equation}\label{sGP}
  - \frac{\hbar^2}{2m} \frac{\triangle \sqrt{n(\x) }}{\sqrt{n(\x)}}+ V(\x) + g' n(\x) = \mu,
\end{equation}
supplemented with the normalization condition $\int\!d\x\, n(\x) = N$.  Here, the coupling strength $g'= \frac{4\pi \hbar^2 a'}{m}$, where $a'$ denotes the $s$-wave
scattering length, and the potential $V(\x)$ is given by Eq.~(\ref{pot}).  In Eq.~\eqref{psi_qc}, $\hat \delta n$ and $\hat \phi$ are the density fluctuation and phase
operators, respectively, which in the Bogoliubov approximation take the following forms
\begin{subequations}
  \label{dn_phi}
  \begin{eqnarray} \label{dn}
    \delta \hat n(\x) &=& \sqrt{n(\x)} \sum_\nu \bigg[f_\nu^-(\x) \hat a_\nu + \mbox{h.c.}\bigg],
    \\ \label{phi}
    \hat \phi(\x) &=& \frac{1}{\sqrt{4 n(\x)}} \sum_\nu  \bigg[- if_\nu^+(\x) \hat a_\nu + \mbox{h.c.}\bigg],
  \end{eqnarray}
\end{subequations}
where ``h.c.'' stands for the hermitian conjugate.
Here, $\hat a_\nu$ are quasiparticle annihilation operator and  $ f_{\nu}^{\pm}$
are mode functions obtained via solution of Bogoliubov-de Gennes equation.
In the case of highly elongated condensates, it turns out that to a very good approximation
the phase operator $\hat \phi(\x)$ depends only on the longitudinal $z$ coordinate~\cite{gerbier}.
In the regime where the thermal fluctuations dominate, 
the modes responsible for phase fluctuation are highly populated, and
we can approximate  the creation and annihilation operators by $c$-numbers~\cite{wasakRaman}, i.e.,  $\hat a_\nu \rightarrow \alpha_\nu$.
This is done together with replacing the quantum average over the thermal state
by an average over a thermal probability distribution, i.e.,
\begin{eqnarray*}
  \langle ... \rangle \rightarrow \langle ... \rangle_{cl} =  \prod_\nu \int \mbox{d}^2 \alpha_\nu \, P(\alpha_\nu) ....,
\end{eqnarray*}
where
\begin{eqnarray}
  \label{distribP}
  P(\alpha_\nu) = \frac{\epsilon_\nu}{k_B T_s \pi} \exp \left(  - \frac{\epsilon_\nu |\alpha_\nu|^2}{k_B T_s} \right).
\end{eqnarray}
As a result, we can replace the operator $\hat\Psi_{qc}$ with a function $\psi_{qc}$:
\begin{equation}\label{decomp}
  \hat \Psi = \psi_{qc}(\x) + \hat \delta (\x)
\end{equation}
where 
\begin{equation}\label{qcpsi}
\psi_{qc}(\x) = \sqrt{n + \delta n(\x)} e^{i\phi(\x)}.
\end{equation}
Here, the functions $\delta n$ and $\phi$ are calculated from Eqs.~\eqref{dn_phi}, with the operators replaced by $c$-numbers drawn from the distribution given
in Eq.~\eqref{distribP}.

We solve the GP equation  (\ref{sGP}) using the Gaussian variational ansatz of the form
\begin{equation}\label{gauss0}
  n(\x) = \frac{N}{\pi^{3/2} \sigma_r^2 \sigma_z} \exp \left( - \frac{x^2+y^2}{\sigma_r^2} - \frac{z^2}{\sigma_z^2}\right).
\end{equation}
The solution, described in details in~\cite{zinnowy}, is:
\begin{eqnarray*}
  && \sigma_r = a_{hor} \left( 1 + \sqrt{\frac{2}{\pi} } \frac{Na'}{\sigma_z} \right)^{1/4},
  \\
  && \left( \frac{\sigma_r \sigma_z}{a_{hoz}^2} \right)^2 = \sqrt{\frac{2}{\pi}} \frac{Na'}{\sigma_z},
\end{eqnarray*}
where $a_{hor} = \sqrt{\hbar/m \omega_r}$ and $a_{hoz} = \sqrt{\hbar/m \omega_z}$.
Substituting the experimental values to the above equations we obtain 
\begin{eqnarray*}
  \sigma_r \simeq 1.7\ \mu\textrm{m},\ \ \ 
  \sigma_z \simeq 0.3\ \textrm{mm}, \ \ \
  \sigma_r/a_{hor} \simeq 1.3.
\end{eqnarray*}
Here we have also used the experimentally determined value for the metastable helium $s$-wave scattering length, i.e., $a'=7.51$~nm~\cite{sl}.

Due to the strong elongation of the cloud and the low temperature only the longitudinal modes are excited.  Therefore, we can approximately treat the system as
quasi-one-dimensional, and define the one-dimensional density:
\begin{equation}\label{n1drow}
  n_{1d}(z) = \int \mbox{d}x \mbox{d}y \, n(\x)
\end{equation} 
and one dimensional interaction constant 
\begin{equation}\label{g1drow}
  g_{1d} = g' \frac{\int \mbox{d} \x \, n^2(\x) }{  \left(\int \mbox{d}\x n(\x) \right)^2}.
\end{equation}
Substituting $n(\x)$ given by Eq.~(\ref{gauss0}) into Eqs.~(\ref{n1drow}) and (\ref{g1drow}), we obtain $n_{1d} \simeq 2\times 10^8 $~atoms/m and $g_{1d}n_{1d}/\hbar\simeq
2\pi \times 2500$~Hz, where $ n_{1d}(z) = n_{1d} \exp(-z^2/\sigma_z^2)$. Using these values, we find the $\gamma = mg_{1d}/\hbar^2 n_{1d}$, which is approximately equal to
$2.5\times 10^{-5}$, i.e., much smaller than unity.  This places us in the weakly interacting regime, and justifies the use of the Bogoliubov method~\cite{gora}.  The
thermal density fluctuations and coherence length of a uniform system take the form:
\begin{equation}  \label{d}
  \frac{\sqrt{\langle \delta n_{1d}^2 \rangle}}{n_{1d}} \simeq \gamma^{1/4} \sqrt{\frac{k_BT_s}{g_{1d}n_{1d}}}, 
  \ \ \ \ \ \ \
  l_\phi = \frac{\hbar^2 n_{1d}}{m k_B T_s}.
\end{equation}
We estimate the density fluctuations and coherence length of our system using the above formula, and obtain
$ \sqrt{\langle \delta n_{1d}^2 \rangle}/ n_{1d} \simeq 0.09$ and  $l_\phi \simeq 120\ \mu\textrm{m}$.
Since the ratio $ \sqrt{\langle \delta n_{1d}^2 \rangle}/ n_{1d} $ is significantly smaller than unity, we neglect the density fluctuations in the further analysis.
We emphasize that $l_\phi$  is smaller than the longitudinal size of the 
cloud but much larger than the transverse size of the cloud $2\sigma_r$.

\section{Quasicondensate collision}\label{sec2}

We show below that the number of scattered atoms is much smaller than the number of atoms in the counter-propagating quasicondensates.  This allows us to neglect the
impact of the scattered atoms on the quasicondensate.  In such a case the evolution of the classical field $\psi_{qc}(\x,t)$ is given by the Gross-Pitaevskii equation
\begin{equation} \label{Gp}
  i \hbar \partial_t \psi_{qc} (\x,t) = \left( - \frac{\hbar^2}{2m} \triangle + g|\psi_{qc}(\x,t) |^2  \right) \psi_{qc}(\x,t).
\end{equation}
Here, the interaction strength is given by $g = 4\pi \hbar^2 a/m$ and $a=5.3$~nm is the scattering length between two atoms in the $m_F=0$ state.  The initial state is:
\begin{equation}
  \psi_{qc}(\x,0) \simeq  \frac{1}{\sqrt{2}} \psi_{qc}(\x) \left( e^{i Q z} + e^{-i Q z} \right)
\end{equation}
and describes a coherent splitting of a single cloud into two components: the function $\psi_{qc}(\x)$ is given by Eq.~(\ref{qcpsi}) and $Q = m v_0/\hbar$.  It is
convenient to factor out the rapidly oscillating phases in time and position, and rewrite the quasicondensate wave function in the following form:
\begin{eqnarray} \nonumber
  \psi_{qc}(\x,t) &=&  \psi_{+Q}(\x,t) \exp\left( iQz  - i \frac{\hbar Q^2}{2m}t\right) 
  \\ \label{psid}
  &+&  \psi_{-Q}(\x,t) \exp\left( - iQz  - i \frac{\hbar Q^2}{2m}t\right) 
\end{eqnarray}
where $\psi_{\pm Q}$ are quasicondensate components moving with mean velocities $\pm v_0 {\bf e}_z$.  In our situation, the momentum width of each of the components is
much smaller than $Q$. Therefore, when determining the momentum density with $\psi_{qc}$, we shall see two separated components. In such a case, the
slowly-varying-envelope approximation can be used~\cite{marek}.  Then, the GP equation can be decomposed, and takes the form
\begin{eqnarray*}
  i \partial_t \psi_{\pm Q} &=& \left(   \mp \frac{\hbar^2}{m} Q \partial_z  -   \frac{\hbar^2}{2m}
  \triangle   \right)\psi_{\pm Q}
  \\
  & & +  g \left( |\psi_{ \pm Q}|^2 + 2|\psi_{\mp Q}|^2  \right)  \psi_{\pm Q}.
\end{eqnarray*}
As we shall see below the time on which the density drops substantially is much smaller than the time needed for the wave-packets to cross each other.  Therefore during
the time important for the collision the motion along the $z$ direction is practically frozen.  This allows us to neglect the terms containing derivatives $ Q \partial_z
$.  As the normalization of each of the wave packet is the same and initially $\psi_{\pm Q}(\x,0) = \frac{1}{\sqrt{2}} \psi_{qc}(\x)$ the above equations turn into a
single one
\begin{equation}\label{SVEA}
  i \partial_t \psi(\x,t) = \left(   -   \frac{\hbar^2}{2m} \triangle  +  \frac{3}{2} g |\psi(\x,t)|^2  \right) \psi(\x,t),
\end{equation}
where the wave function
\begin{equation} \label{rowpsi}
  \frac{1}{\sqrt{2}} \psi = \psi_{+Q} = \psi_{-Q},
\end{equation}
with the normalization condition $\int\!d\x \, |\psi(\x,t)|^2 = N$ and initial condition $\psi(\x,0) = \psi_{qc}(\x)$.

We have found above that the quasicondensate coherence length is much larger than the transverse size of the cloud. Therefore, we expect that the expansion of the cloud
in the transverse directions is practically not affected by its quasicondensate nature.  Thus, it is instructive to consider expansion of the condensate, i.e.,
the solution $\psi_c(\x,t)$ of the above Eq.~\eqref{SVEA} with the initial condition $\psi_c(\x,0) = \sqrt{n(\x)}$.  We approach the problem with an approximate variational
gaussian ansatz, described in \cite{zinnowy}, and obtain:
\begin{eqnarray} \nonumber
  && \psi_c(\x,t) \simeq \sqrt{\frac{N}{\pi^{3/2} \sigma_z(t) \sigma_r^2(t)}} \exp \left(  - \frac{z^2}{2\sigma_z^2(t)} - i b_z(t) z^2
  \right)
  \\ \label{psicw}
  && \times \exp \left( - \frac{x^2+y^2}{2\sigma_r^2(t)}
  \left( 1 - i \tilde \omega t \frac{\sigma_r^2}{\tilde a_{hor}^2} \right)- i \varphi(t) \right),
\end{eqnarray}
where 
\begin{eqnarray*}
  && \sigma_r^2(t) = \sigma_r^2(1+\tilde \omega^2 t^2),
  \ \ \ \ \
  \frac{\tilde \omega^2}{\omega_r^2} = \frac{1 + \sqrt{\frac{2}{\pi}} \frac{3}{2} \frac{Na}{\sigma_z}}{
    1 + \sqrt{\frac{2}{\pi}} \frac{Na'}{\sigma_z}}, \ \ \ \ 
  \\
  && \varphi(t) = \left( \frac{7}{4} \frac{\sigma_r^2}{\tilde a_{hor}^2} - \frac{3}{4} \frac{\tilde a_{hor}^2}{\sigma_r^2} \right)
  \arctan (\tilde \omega t),
\end{eqnarray*}
and $\tilde a_{hor} = \sqrt{\hbar/m\tilde \omega}$.  With the parameters of the experiment, we obtain $\tilde \omega \simeq 1.02 \omega_r$.  This gives us the
characteristic time of expansion expansion $\tau_{ex } = 1/\tilde \omega = 104\ \mu\mathrm{s}$.  This time allows us to estimate the change of the phase caused by thermal
fluctuations.  To this end, we consider a one-dimensional uniform system and calculate $ \Delta\phi \equiv \sqrt{\langle (\phi(z=0,\tau_{ex})-\phi(z=0,0))^2 \rangle} $;
the details of the evaluation are presented in Appendix~\ref{Dodatekfaza}.  There, we find that $\Delta\phi \simeq 0.14 $, and since it is significantly smaller than
unity, we expect that the change of phase during the collision, caused by thermal fluctuations, is negligible.  Thus, we assume that
\begin{equation}\label{psi11}
  \psi_{+Q}(\x,t) = \psi_{-Q}(\x,t) = \sqrt{2}\psi_c(\x,t) e^{i \phi(z)}.
\end{equation}

Additionally we find the expansion time $\tau_{ex}$ to be much smaller than the time needed for the quasicondensates to cross each other, $2\sigma_z/v_0 \simeq
6$~ms. This justifies the use of the approximation which leads to Eq.~(\ref{SVEA}).

We now make one more approximation. From the formulas presented in \cite{zinnowy}, we find that the phase gradient $ b_z(\tau_c) \sigma_z$ is much smaller than
$\partial_z \phi(z)$ and thus can be neglected. Additionally, we find that $\sigma_z(\tau_{ex}) \simeq \sigma_z$.  As a result, in what follows, we use $\psi_c(\x,t)$
given by Eq.~(\ref{psicw}) with $\sigma_z(t) \simeq \sigma_z$ and $b_z(t) \simeq 0$. For simplicity of the notation, we express the variational ansatz, given by
Eq.~(\ref{psicw}), decoupling its $z$ dependence:
\begin{eqnarray} \label{psicN}
  \psi_c(\x,t) &=& \psi_\rho(x,y,t) \exp \left( -\frac{z^2}{2\sigma_z^2}  \right),
  \\ \nonumber
  \psi_\rho(x,y,t) &=& \sqrt{n_0} \frac{\sigma_r}{\sigma_r(t)}   \exp \left( - i \varphi(t) \right),
  \\ \nonumber
  &\times&
  \exp \left( - \frac{x^2+y^2}{2\sigma_r^2(t)}
  \left( 1 - i \tilde \omega t \frac{\sigma_r^2}{\tilde a_{hor}^2} \right) \right),
\end{eqnarray}
where $n_0 = {N}/({\pi^{3/2} \sigma_z \sigma_r^2})$.

\section{The scattered atoms} 
\label{sec3}

We now turn our attention to the description of the scattering process.  As in the experiment, we consider the scattered atoms with velocities restricted to $ \frac{\pi}{3} <
\theta < \frac{2\pi}{3}$, where $\theta$ is the angle between velocity of the clouds and the $z$ axis.  Thus, the average distance traveled by the scattered atom within
the cloud is approximately given by $2 \sigma_r \simeq 3.4\ \mu\textrm{m}$.  This distance is much smaller than the mean free path equal to $ 1/(8\pi {a'}^2 n_0) \simeq
66\ \mu\mathrm{m} $.  As a result, the system is in the collisionless regime and the use of the Bogoliubov approximation to treat the scattered atoms is adequate. The
field operator $\hat \delta$, describing the scattered atoms, undergoes the time evolution given by
\begin{eqnarray}\label{glowne}
  i \hbar \partial_t \hat \delta (\x ,t) =  H_0 (\x ,t) \hat \delta (\x ,t) +  B (\x ,t) \hat \delta^\dagger (\x ,t),
\end{eqnarray}
where:
\begin{eqnarray}\label{H0}
  H_0(\x ,t)  &=& - \frac{\hbar^2}{2m} \triangle + 2 g |\psi_{qc} (\x,t)|^2,
  \\ \nonumber
  \\ \label{B0}
  B(\x,t) &=& g \psi_{qc}^2 (\x,t).
\end{eqnarray}
We assume that the initial state of the noncondensed particles is vacuum~\cite{dopiska}, i.e.,
\begin{equation}\label{stan}
  \hat \delta(\x,0)|0 \rangle = 0.
\end{equation}
Since the scattered atoms for chosen $\theta$ do not overlap with the quasicondensates,
the pair correlation function is given by the formula
\begin{equation} \label{G2def}
  G^{(2)}\left( \x_1,\x_2,T \right) = \langle \langle \hat \delta^\dagger (\x_1 ,T) \hat \delta^\dagger (\x_2 ,T)
  \hat \delta (\x_2 ,T) \hat \delta (\x_1 ,T)\rangle \rangle_{cl},
\end{equation}
where we deal with quantum average over degrees of freedom described by the $\hat \delta$ operator and the classical average
over quasicondensate modes.
In the above, $T \simeq 0.3 \, \mbox{s}$ is the time it takes the atoms to reach the MCP located
$46\ \mathrm{cm}$ below the trapped cloud.

As the equation of motion for the field operator $\hat \delta $, given by Eq.~(\ref{glowne}), is linear,
and the quantum state is the vacuum, the Wick theorem can be applied.
As a result, the quantum average can be evaluated and it reads
\begin{eqnarray} \nonumber
  &&  \langle \hat \delta^\dagger (\x_1 ,T) \hat \delta^\dagger (\x_2 ,T)
  \hat \delta (\x_2 ,T) \hat \delta (\x_1 ,T)\rangle
  \\ \nonumber
  && = G^{(1)} \left( \x_1,\x_1,T \right)G^{(1)} \left( \x_2,\x_2,T \right)
  +  \left|G^{(1)} \left( \x_1,\x_2,T \right) \right|^2
  \\ \label{G2n}
  && + \left| M \left( \x_1,\x_2,T \right)  \right|^2,
\end{eqnarray}
where
\begin{equation}\label{Mdef}
  M\left( \x_1,\x_2,T \right) \equiv  \langle \hat \delta (\x_1 ,T)
  \hat \delta (\x_2 ,T)\rangle
\end{equation}
is called the anomalous density and
\begin{equation}\label{G1}
  G^{(1)} \left( \x_1,\x_2,T \right) \equiv \langle \hat \delta^\dagger (\x_1 ,T)
  \hat \delta (\x_2 ,T)\rangle.
\end{equation}
is a single particle correlation function of the scattered atoms in a single realization of the quasicondensate field.

In Eq.~(\ref{G2n}), two terms are responsible for the correlations: $| M \left( \x_1,\x_2,T \right)|^2$ and
$|G^{(1)} \left( \x_1,\x_2,T \right)|^2$. The term
$G^{(1)} \left( \x_1,\x_1,T \right)G^{(1)} \left( \x_2,\x_2,T \right)$ is a
product of single particle densities and represents uncorrelated particles.
In the next sections, we show that the terms $| M \left( \x_1,\x_2,T \right)|^2$ and $|G^{(1)} \left( \x_1,\x_2,T \right)|^2$
represent the correlation of particles with opposite and collinear velocities, respectively.
The appearance of correlations of particles with opposite velocities (which we shall call the ``opposite" correlation)
is due to the fact that particles are scattered in pairs of opposite momenta.
On the other hand, the correlation of particles with collinear velocities (which we shall call the ``local" correlation)
is a bosonic bunching effect. Therefore, we introduce notation 
$G_{op}^{(2)} = |M|^2$ and $G_{loc}^{(2)} = |G^{(1)}|^2$.

To calculate the correlations, we solve the Heisenberg equation of motion, Eq.~(\ref{glowne}), with the  perturbation approach~\cite{zinnowy}.
Then, the formula for the anomalous density reads
\begin{eqnarray}\label{M}
  M(\x_1,\x_2,T)
  &=&  \frac{1}{i \hbar} \int_0^T \mbox{d} t \int \mbox{d}\x \, K(\x_1 ,T; \x,t)
  \\ \nonumber
  & & \ \ \ \ \ \ \ \ \ \ \ \ \ \ \ \ \times K(\x_2 ,T; \x,t)   B (\x ,t),
\end{eqnarray}
where $K(\x_1,t_1;\x_2,t_2)$ is a single body propagator of the Hamiltonian from Eq.~(\ref{H0}).
Additionally, one obtains a simple relation between one body correlation function and the anomalous density
\begin{equation}\label{pertG1}
  G^{(1)} \left( \x_1,\x_2,T \right) =\int \mbox{d} \x \, M^*\left( \x_1,\x,T \right)M\left( \x,\x_2,T \right).
\end{equation}

In Ref.~\cite{zinnowy}, we analyzed the case of two colliding condensates.
There, we employed a semiclassical approximation for the propagator $K$ that 
leads to the following formula for the anomalous density in the momentum space:
\begin{eqnarray} \nonumber
  && M(\KK, \DK )
  = \frac{1}{i\hbar (2\pi)^3}   \int_0^\infty \mbox{d} t \int \mbox{d} \x \,
  \\ \label{Mk2}
  &&
  e^{  - i \DK \x + i \frac{\hbar}{m} \left( K^2 + \frac{\Delta K^2}{4}  \right)  t } \widetilde B({\bf e}_\KK,\x,t),
\end{eqnarray}
where
\begin{subequations}
  \label{semi-formulas}
  \begin{eqnarray}
    \label{Bmf}
    \widetilde B({\bf e}_\KK,\x,t) &=& B(\x,t )\exp \left(- i  \Phi(\x,{\bf e}_{\KK},t) \right),
    \\ \label{MFPhi}
    \Phi(\x,{\bf e}_{\KK},t) &=&  \frac{m}{\hbar^2 Q} \int^\infty_{-\infty} \mbox{d} s \, V_{en}(\x  + s  {\bf e}_{\KK},t),
    \\ \label{Ven111}
    V_{en}(\x,t) &=& 2g \left( |\psi_{+ Q} (\x,t)|^2 + |\psi_{- Q} (\x,t)|^2 \right),
  \end{eqnarray}
\end{subequations}
and ${\bf e}_{\KK} = \frac{\KK}{K}$. Here, $\psi_{\pm Q}$ are the two counter propagating components of the condensate.  The mean-field
interaction $2g|\psi|^2$, present in the Hamiltonian $H_0$ in Eq.~(\ref{H0}), is represented by the potential $V_{en}$ and enters the formulas through the phase $\Phi$
given by Eq.~(\ref{MFPhi}).  If we omitted the mean-field interaction in the Hamiltonian $H_0$, the phase $\Phi$ would be zero.

The formulas \eqref{semi-formulas} were derived for two counter propagating condensates.  In Appendix~\ref{app2}, we show that they also work in the case of
quasicondensates.  Inserting Eqs.~(\ref{rowpsi}) and (\ref{psi11}) into Eq.~(\ref{Ven111}), we arrive at
\begin{equation} \label{Ven}
  V_{en}(\x,t) = 2g |\psi_c (\x,t)|^2.
\end{equation}

We now focus on the expression for $B(\x,t)$.  According to Eq.~(\ref{B0}), we have $B(\x,t) = g\psi_{qc}^2(\x,t)$.  Inserting $\psi_{qc}$, given by Eq.~(\ref{psid}), we
arrive at an expression with three terms.  However, among them there is only one responsible for the scattering of atoms from collision between $\pm Q$ components.  As we
are interested only in this process, we neglect the others arriving at
\begin{eqnarray} \nonumber
  B(\x,t) &\simeq& 2g \psi_{+Q} \psi_{-Q}(\x,t) \exp \left( - i \frac{\hbar Q^2}{m} t \right)
  \\ \label{Ba}
  &=&  g\psi_c^2 (\x,t) \exp \left( 2i \phi(z) - i \frac{\hbar Q^2}{m} t \right),
\end{eqnarray}
where we used Eqs.~(\ref{rowpsi}) and (\ref{psi11}).

Finally, we note that the anomalous density in real space is connected with the momentum space representation via the standard free propagator formula
\begin{eqnarray} \label{Mrealspace}
  && M(\X,\DX,T) =  \frac{1}{(2\pi)^3}
  \int \!\!d \KK\, d \DK \, e^{ i 2 \KK \X   } \times
  \\ \nonumber
  && \quad \quad \times e^{ i \frac{\DK \DX}{2}- i\frac{\hbar }{m} \left( K^2 + \frac{\Delta K^2}{4} \right) T  }
  M(\KK,\DK),
\end{eqnarray}
where $\X = ({\x_1-\x_2})/{2}$ and $\DX = \x_1+\x_2 $.

\section{The $G^{(1)}$ and $G^{(2)}_{loc}$ functions} \label{sec4}

We have 
\begin{equation}\label{G2locdef}
  G^{(2)}_{loc}(\x_1,\x_2,T) = \langle  \left| G^{(1)}(\x_1,\x_2,T)   \right|^2 \rangle_{cl}.
\end{equation}
In order to calculate $G^{(1)}$, it is convenient to use the source function $f$ description introduced in Ref.~\cite{zinnowy}.
The motivation of introducing such a function comes from classical physics where the single particle phase-space density
$W(\x,\K,T)$ of the  particles emitted from a source takes the form:
\begin{eqnarray}\label{Wc}
  W(\x,\K,T) = \int_0^T \!\!dt \, f\!\left(\x - \frac{\hbar \K}{m}(T-t),\K, t  \right),
\end{eqnarray}
where $f(\x,\K,t)$ is the source function describing a density of particles emitted by the source 
in position $\x$ with momentum $\hbar \K$ and in time $t$.
Now, we use the above equation to define $f$ in a quantum system by assuming that $W$ is the Wigner distribution.
Then, using the relation
\begin{eqnarray*}
  G^{(1)}\left( \x + \frac{\Delta \x}{2}, \x - \frac{\Delta \x}{2}, T\right) = \int \mbox{d} \K \, e^{-i\K \Delta \x} W(\x,\K;T),
\end{eqnarray*}
we arrive at
\begin{eqnarray*}
  \widetilde G^{(1)}\left( \x ,\Delta \x, T\right) &=& \int_0^T \mbox{d} t  \int \mbox{d} \K \, e^{  - i \K \dx }
  \\
  & &   f\left(\x - \frac{\hbar \K}{m}(T-t),\K, t  \right),
\end{eqnarray*}
where we denoted $G^{(1)}\left( \x + \frac{\Delta \x}{2}, \x - \frac{\Delta \x}{2}, T\right) $ by  $ \widetilde G^{(1)}\left( \x ,\Delta \x, T\right)  $. 
Moreover, introducing  new variables: $\x' = \x - \frac{\hbar}{m} \K (T-t)$, $\K_0 = \frac{m\x}{\hbar T}$ and $\dk_0 = \frac{m \dx}{\hbar T}$, we obtain
\begin{eqnarray} \nonumber
  && \widetilde G^{(1)}\left( \x ,\dx, T\right) = \int_0^T \mbox{d} t  \int \mbox{d} \x' \, \left(\frac{m}{\hbar(T-t)} \right)^3 \\ 
  &&
  e^{  - i \frac{1}{1-\frac{t}{T} } \left(\frac{\hbar}{m} \K_0 T  - \x' \right)  \dk_0   } 
  f\!\left(\x', \!\frac{ \K_0 - \frac{m \x'}{\hbar T} }{1-\frac{t}{T}} , t\!  \right).
  \label{G1f}
\end{eqnarray}

We now analyze the source function. In Ref.~\cite{zinnowy}, we showed that the source function is equal to:
\begin{eqnarray} \nonumber
  f(\x , \K, t ) &=&   \frac{2}{(2\pi)^3\hbar^2} \int_{-t}^{t} \!d\Delta t  \, e^{ - i \frac{2\hbar k^2}{m} \Delta t }\times
  \\ \label{sourcef}
  & \times &  B_p^*(\x,\K,t, -\Delta t) B_p(\x,\K,t, \Delta t),
\end{eqnarray}
where
\begin{eqnarray}\nonumber
  B_p(\x,\K,t,\Delta t) &=& \int \mbox{d} \Delta \x \, K_f(\Delta \x,\Delta t) \times
  \\ \label{Bpwzor}
  & \times &  \!\widetilde B\left({\bf e}_\K,\x + \Delta \x + \frac{\hbar \K \Delta t}{m}, t -\Delta t\!\right),
\end{eqnarray}
and $K_f$ denotes the free propagator. We now analyze the above formulas in the way analogous to that presented in Ref.~\cite{zinnowy}.
It is crucial to note that $\widetilde B$ vanishes if  $|x+\Delta x + \frac{\hbar k_x \Delta t}{m}|$ is larger than $\sigma_r$.
As  $\K=(k\sin \theta,0,k\cos\theta)$ and $k \simeq Q$ with $\sin \theta > \sqrt{3}/2$, we
find that $|\Delta t| < \Delta t_0 = \frac{2\sigma_r}{v_0 \sqrt{3}} \simeq 20\ \mu\textrm{s}$. 
We note that $\Delta t_0$ is the time the scattered particle leaves the cloud. 
The characteristic distance $\Delta_0 = \sqrt{\hbar \Delta t_0/m} \simeq 0.56\ \mu\mathrm{m}$ present in the free propagator
$K_f(x,t) \propto \exp \left( \frac{i x^2}{2 \Delta_0^2 t/t_0} \right) $
is more than three times smaller than $\sigma_r$. 
On the distance $\Delta_0$ and in time $\Delta t_0$, the change of the wave function $\psi_c$ and
the phase $\Phi$ is not crucial and can be neglected.
As a result, from Eqs.~(\ref{Bmf}), (\ref{Ba}) and (\ref{Bpwzor}), we obtain that  
\begin{equation}\label{Bpw2}
  B_p(\x,\K,t,\Delta t) \simeq  
  \widetilde B\left({\bf e}_\K,\x + \frac{\hbar \K \Delta t}{m}, t\right)
  \exp \left( i \frac{\hbar Q^2}{m} \Delta t \right).
\end{equation}

In the formula describing the source function $f$, and given by Eq.~(\ref{sourcef}),
we notice the presence of the term $B_p^*(\x,\K,t, -\Delta t) B_p(\x,\K,t, \Delta t)$.
According to the above equation, it is now equal to 
\begin{eqnarray} \nonumber
  && \widetilde B^*\left({\bf e}_\K,\x - \frac{\hbar \K \Delta t}{m}, t\right)
  \widetilde B\left({\bf e}_\K,\x + \frac{\hbar \K \Delta t}{m}, t\right)\times
  \\
  && \label{narazie}
  \times \exp \left( i \frac{2 \hbar Q^2}{m} \Delta t \right)
\end{eqnarray}
Substituting here  $\widetilde B$, given by Eqs.~(\ref{Bmf}) and (\ref{Ba}), and using the fact that the phase $\Phi$ is constant along ${\bf e}_\K$
direction, i.e., $\Phi(\x-s{\bf e}_\K,{\bf e}_\K,t)=\Phi(\x,{\bf e}_{\K},t)$, we find a cancellation
of the phase $\Phi$ in Eq.~(\ref{narazie}).
Additionally, $\frac{\hbar k_z \Delta t}{m} $ is maximally equal to $\sigma_r$.
On that distance, we can neglect the change of the phase $\phi(z)$ and, as a result, 
we obtain a cancellation of the phase $\phi$ in Eq.~(\ref{narazie}).  As a result,
we find that the quantity present in Eq.~(\ref{narazie}) is equal to
\begin{equation}  \label{narazie2}
  g^2
  {\psi_c^2}^* \left(\x - \frac{\hbar \K \Delta t}{m}, t\right)
  \psi_c^2 \left(\x + \frac{\hbar \K \Delta t}{m}, t\right).
\end{equation}
Consequently, we find that the source function $f$ does not depend on the phases $\Phi$ and $\phi$. Due to the relation given by Eqs.~(\ref{G1f}) and (\ref{G2locdef}),
the same applies to a single particle correlation function $G^{(1)}$ and local part of pair correlation function $G^{(2)}_{loc}$.  It means that these functions are not
affected by the presence of the quasicondensate nor by the interaction of the scattered atoms with the atoms of the colliding clouds. Therefore, we can omit the bracket
$\langle \ldots \rangle_{cl}$ present in Eq.~(\ref{G2locdef}), and arrive at
\begin{equation}
  G^{(2)}_{loc}(\x_1,\x_2,T) = \left|G^{(1)}(\x_1,\x_2,T) \right|^2.
\end{equation}

We now continue with the analysis of the source function $f$.  From Eqs.~(\ref{sourcef}), (\ref{Bpw2}) and (\ref{narazie2}), we find that
\begin{eqnarray} \nonumber
  f(\x , \K, t ) &=&   \frac{2}{(2\pi)^3\hbar^2} \int_{-t}^{t} \mbox{d} \Delta t  \, 
  \exp \left( - i \frac{2\hbar (k^2-Q^2)}{m} \Delta t \right)\times
  \\ \nonumber
  & & g^2 {\psi_c^*}^2 \left(\x - \frac{\hbar \K \Delta t}{m},t \right)
     {\psi_c}^2 \left(\x + \frac{\hbar \K \Delta t}{m},t \right).
\end{eqnarray} 
To proceed further, we make a series of approximations.
First, $\frac{\hbar k_z \Delta t}{m} $ is maximally equal to $\sigma_r$. 
As $\sigma_r \ll \sigma_z$, we neglect the change of $\psi_c$ on that distance.
Second,  $\Delta t_0$ is significantly smaller than the collision time. Therefore, we  approximate
$\int_{-t}^t \approx \int_{-\infty}^\infty$.
Third, as $|k-Q| \ll Q$, we approximate $k^2-Q^2 \simeq 2 Q \delta k$, where $\delta k = k - Q$,
and additionally approximate $\frac{\hbar \K \Delta t}{m} \simeq \frac{\hbar Q \Delta t}{m} e_\K $.
As a result, the above formula takes the form
\begin{eqnarray} 
  \nonumber
  &&  f(\x , \K, t ) =   \frac{m {g'}^2}{ 4\pi^3\hbar^3 Q}  \int \mbox{d} \delta r  \, 
  \exp \left( - i 4 \delta k \delta r  \right)\times
  \\ \nonumber
  & & \quad  {\psi_c^*}^2 \left(x - \delta r \sin \theta,y,z,t \right)
     {\psi_c}^2 \left(x+\delta r \sin \theta,y,z,t \right),
\end{eqnarray} 
where we introduced $\delta r = \frac{\hbar Q \Delta t}{m}$ and used ${\bf e}_\K = (\sin \theta,0,\cos \theta)$.
Inserting into the above $\psi_c$ given by Eq.~(\ref{psicN}) and performing the integral over $\delta  r$,
we obtain
\begin{eqnarray} \nonumber
  f(\x , \K, t ) &=&  \frac{D_1}{(1+\tilde \omega^2 t^2)^{3/2}\sin \theta}
  e^{ - 2\frac{z^2}{\sigma_z^2}    - 2\frac{x^2+y^2}{\sigma_r^2(t)} } \times
  \\ \label{fwynik}
  &\times&  \exp\!\! \left[  - 2\left( \frac{\delta k \sigma_r(t)}{\sin \theta} 
    - \frac{x}{\sigma_r(t)} \tilde \omega t \frac{\sigma_r^2}{\tilde a_{hor}^2} \right)^2
    \right]\!,
\end{eqnarray}
where $D_1 = \frac{\sqrt{\pi} m g^2 n_0^2 \sigma_r }{4\sqrt{2}\pi^3\hbar^3 Q } = \frac{2 \sqrt{2}}{\sqrt{\pi}} n_0^2 a^2 \sigma_r \tilde a_{hor}^2 \frac{\tilde \omega}{Q}
$.  This formula gives an intuitive understanding of the source function.  One might wonder if the above function can be derived within a classical model, where one
considers a collision of two clouds of atoms. It turns out that it is indeed the case, when one takes the Wigner function as the phase-space distribution of the colliding
clouds~\cite{zinnowy}.

We find that, if $\sigma_r \ll \sigma_z $, $\frac{m \sigma_z}{\hbar T} \ll \frac{1}{\sigma_r}$ and $\tilde \omega T \gg 1$, the formula for the $\tilde G^{(1)}$ function,
given by Eq.~(\ref{G1f}), can be approximated as:
\begin{eqnarray} \nonumber
  && \widetilde G^{(1)}\left( \frac{\hbar\K_0}{m } T , \frac{\hbar\dk_0}{m}T, T\right) \simeq 
  \left(\frac{m}{\hbar T} \right)^3
  \exp \left(  - i  \frac{\hbar}{m} \K_0  \dk_0 T   \right)
  \\ \label{G1rrr}
  &&
  \int_0^T \mbox{d} t  \int \mbox{d} \x' \,
  \exp \left(  i \dk_0\left(\x' -\frac{\hbar}{m} \K_0 t  \right)     \right)  f\left(\x', \K_0, t  \right).
\end{eqnarray}
With the experimental parameters, we verify that the conditions given above are satisfied, and this formula 
can be applied to our system.

Now, we proceed to the analysis of the density of scattered particles.
From Eq.~(\ref{G1rrr}),  we obtain
\begin{eqnarray} \nonumber
  \varrho(\K) &=&  \langle \widetilde G^{(1)} \left( \frac{\hbar\K}{m } T , 0, T\right) \rangle_{cl}
  = \widetilde G^{(1)} \left( \frac{\hbar\K}{m } T , 0, T\right)
  \\ \nonumber
  & \simeq& 
  \left(\frac{m}{\hbar T} \right)^3 \int_0^\infty \mbox{d} t  \int \mbox{d} \x \, f\left(\x, \K, t  \right).
\end{eqnarray}
Inserting $f$ from Eq.~(\ref{fwynik}) into the above, we obtain  
\begin{equation}\label{rho111}
  \varrho(\K) = \left(\frac{m}{\hbar T} \right)^3\frac{1}{\sin \theta} h \left( \frac{\delta k}{\sin \theta} \right),
\end{equation}
where $h$ is given by 
\begin{eqnarray} \nonumber
  h(\delta k)   &=&   D_2
  \int_0^\infty \frac{\mbox{d} \tilde t}{ \sqrt{\left(1 +    \tilde  t^2 \frac{\sigma_r^4}{\tilde a_{hor}^4 } \right)
      \left( 1+\tilde  t^2 \right)}}
  \\ \label{h111}
  & & \exp \left(  - 2  \frac{\delta k^2 \sigma_r^2 (1+\tilde  t^2)}{
    \left(1 +    \tilde t^2 \frac{\sigma_r^4}{\tilde a_{hor}^4 }  \right)  }   \right),
\end{eqnarray}
and the parameters $D_2 = D_1   \left(\frac{\pi}{2} \right)^{3/2} \sigma_r^2\sigma_z/\tilde \omega
= \frac{\pi}{ Q} n_0^2 a^2 \sigma_r^3 \tilde a_{hor}^2 \sigma_z$ and $\tilde t = \tilde \omega t$.

Now we turn our attention to the normalized two-body correlation function
measured in the experiment, and defined as
\begin{eqnarray} \label{g2}
  &&
  g^{(2)}_{loc}(\Delta r,\Delta z) -1 =  \\
  &&
  = \frac{\int \mbox{d} \tilde \dx \ w(\dx -\tilde \dx) \int_V \mbox{d} \x \, G^{(2)}_{loc}(\x_1,\x_2,T) }{  \int \mbox{d} \tilde \dx \ w( \dx -\tilde \dx )
    \int_V \mbox{d} \x \, G^{(1)}(\x_1,\x_1,T) G^{(1)}(\x_2,\x_2,T) }, \nonumber
\end{eqnarray}
where $\x_1=\x + \tilde \dx/2 $, $\x_2 = \x - \tilde \dx /2 $
and $V$ denotes a volume where the spherical angles $\x =r (\sin \theta \cos \phi, \sin \theta \sin \phi,\cos\theta) $
are bounded by $ \frac{\pi}{3}  < \theta < \frac{2\pi}{3}$.
The function  $w(\x)$ is given by
\begin{equation}\label{w}
  w(\x) = \frac{1}{(2\pi)^{3/2} \sigma_{zd} \sigma_{rd}^2 } e^{- \frac{x^2+y^2}{2 \sigma_{rd}^2} - \frac{z^2}{2\sigma_{zr}^2} },
\end{equation}
and describes the detector resolution for two-particle detection.  The transverse resolution is known to be $\sigma_{rd} = 350\ \mu\mathrm{m}$~\cite{HBT0}.  The vertical
resolution $\sigma_{zd}$ has never been precisely measured, but from Ref.~\cite{tunableParis}, we can place an upper limit $\sigma_{zd} < 60 \ \mu\mathrm{m}$.  Due
to the cylindrical symmetry of the system, $g^{(2)}_{loc}$ depends only on $\Delta r = \sqrt{\Delta x^2 + \Delta y^2 }$ and $\Delta z$.  Therefore, we denote
it as $g^{(2)}_{loc}(\Delta r, \Delta z)$.

\section{The $G^{(2)}_{op}$ function}\label{sec5}

In this section, we consider the object
\begin{eqnarray}\nonumber
  G^{(2)}_{op} \left(\DX,T\right) &=& \int \mbox{d} \X \, G^{(2)}_{op} \left( \X,\DX,T\right)
  \\ \label{G2bbav}
  &=& \int \mbox{d} \X \, \langle |M(\X,\DX,T)|^2 \rangle_{cl},
\end{eqnarray}
where 
\begin{equation}\label{G2bb}
  G^{(2)}_{op} \left( \X,\DX,T\right) = \langle |M(\X,\DX,T)|^2 \rangle_{cl},
\end{equation}
with $M$ given by Eq.~(\ref{Mrealspace}).  The analysis of $G^{(2)}_{op}$ is performed in Appendix~\ref{G2bbApp}.  Below, we summarize the main
results of our analysis.

The crucial step made in the calculation of $G^{(2)}_{op} \left(\DX,T\right)$ is the use of $\psi_c$ given by the variational approximation
and presented in Eq.~(\ref{psicN}).  Due to a Gaussian form of the ansatz and the cylindrical symmetry of the system, $B(\x,t)$, given by Eq.~(\ref{Ba}), decomposes into
$B(\x,t) = B_\rho(\sqrt{x^2+y^2},t)B_z(z)$.  Additionally, the ansatz allows an analytical formula for the phase $\Phi(\x,{\bf e}_\KK,t)$ in
Eq.~(\ref{MFPhi}).

It turns out that the gradient of the quasicondensate phase $\partial_z \phi(z)$ is much larger than $\partial_z \Phi$. This leads to an approximation where $\Phi(\x,{\bf
  e}_\KK,t)$ is replaced by its value averaged over $z$, denoted by $\widetilde \Phi(x,y,{\bf e}_\KK,t)$.  As a result, $\widetilde B({\bf e}_\KK,\x,t)$, given by
Eq.~(\ref{Bmf}), decomposes into $\widetilde B({\bf e}_\KK,\x,t) = \widetilde B_\rho({\bf e}_\KK,\sqrt{x^2+y^2},t) B_z(z)$.  This again allows a decomposition of the
$G^{(2)}_{op} \left(\DX,T\right) $:
\begin{eqnarray} \nonumber
  G^{(2)}_{op} \left(\DX,T\right)  \simeq   \int_0^\infty \mbox{d} t \int \mbox{d} \Omega_\KK  
  && G^{(2)}_z(\Delta Z,T) 
  \\ \label{G2bbG}
  && \times
  G^{(2)}_\rho(\Delta R, {\bf e}_\KK, t),
\end{eqnarray}
where $\Delta R = \sqrt{\Delta X^2 + \Delta Y^2}$.  Note, $ G^{(2)}_z $ does not depend on ${\bf e}_\KK$ or $t$, which follows directly from the lack of this dependence
in the $B_z$ function.

For non-interacting particles and after sufficiently long expansion time,
the cloud has the shape given by its velocity distribution.
As a consequence, all the spatial correlations will be given by their momentum counterparts.
Due to a large free-fall time,  we might
expect that this situation occurs also  in our system,
i.e., that $G_{op}^{(2)} \left( \X,\DX,T\right)$ in position space is proportional to 
$G_{op}^{(2)}$ in momentum space.
Indeed, in Appendix~\ref{ap33} we show that this is indeed the case for the $\Delta X$, $\Delta Y$
and $\X$ variables. 
However, for the variable $\Delta Z$, the time $T$ is not sufficiently long
to obtain the momentum space result.
Instead, we arrive at  
\begin{eqnarray*}
  G^{(2)}_z(\Delta Z,T) = \int   \mbox{d} K_z \, G^{(2)}_z\left(\Delta Z - \frac{\hbar T}{m} K_z ,0 \right)
  \rho(K_z),
\end{eqnarray*}
where $\rho(K_z) = \frac{2\pi l_\phi}{1+(K_z l_\phi /2)^2}$ is the velocity distribution directly related to the quasicondensate velocity distribution and $
G^{(2)}_z(\Delta Z,0) = \exp \left( -\frac{\Delta Z^2}{2\sigma_z^2} \right)$ is the initial correlation function.  Here, we clearly notice the classical formula for the
propagation of the initial correlation.  For very small $T $, the correlation function is simply given by the spatial dependence of the cloud $ \exp \left( -
\frac{\Delta Z^2}{2 \sigma_z^2} \right)$.  For $T \rightarrow \infty$, it will be given by the velocity distribution $\rho$.  For the experimental parameters, we have
$\frac{\hbar T }{m l_\phi \sigma_z} \simeq 0.14$, which effectively leads to small broadening of the initial spatial distribution.  The gaussian function fitted to the
distribution takes the form
\begin{equation} \label{Mzr}
  G^{(2)}_z(\Delta Z,T) \simeq (2\pi)^2 \frac{\sigma_z}{ \tilde \sigma_z} \exp \left( - \frac{\Delta Z^2}{2 \tilde \sigma_z^2} \right)
\end{equation}
where $\tilde \sigma_z \simeq 1.17 \sigma_z$.  Due to smallness of the parameter $\frac{\hbar T }{m l_\phi \sigma_z} $, we notice practically negligible impact of the
presence of the quasicondensate on $G^{(2)}_{op} \left(\DX,T\right) $.

The $\Delta R$ correlations are related to the $\psi_\rho$ function [see Eq.~(\ref{psicN})], describing the expansion of the wave function in transverse directions,
and to the phase $\widetilde \Phi$. These correlations are not affected by the presence of the quasicondensate. The formulas for $G^{(2)}_\rho(\Delta R,
{\bf e}_\KK, t) $ are presented in Appendix~\ref{G2bbApp}.

Finally, we note that the pair correlation function measured in the experiment takes the form
\begin{widetext}
  \begin{equation} \label{g2bb}
    g^{(2)}_{op}(\Delta R, \Delta Z) =\frac{\int \mbox{d} \tilde \DX \ w(\DX -\tilde \DX)  
      G^{(2)}_{op}(\DX,T) }{   \int \mbox{d} \tilde \DX \ w( \DX -\tilde \DX )
      \int_V \mbox{d} \X \, G^{(1)}(\x_1,\x_1,T) G^{(1)}(\x_2,\x_2,T)   }
  \end{equation}
\end{widetext}
where $\x_1= \X + \tilde \DX/2 $, $\x_2 = -\X + \tilde \DX /2 $, $\Delta R = \sqrt{\Delta X^2 + \Delta Y^2}$, and $V$ denotes a volume where the spherical angles $\x =r
(\sin \theta \cos \phi, \sin \theta \sin \phi,\cos\theta) $ are bounded by $ \frac{\pi}{3} < \theta < \frac{2\pi}{3}$. Therefore, we have all the ingredients to calculate
the local and oposite correlations.


\section{Results} \label{sec6}

To make a comparison between theory and experiment, we first examine
$g^{(2)}_{op}(0,0)$, the amplitude of the opposite correlation, given by Eq.~(\ref{g2bb}), and
using  Eqs.~(\ref{G2bbG}), (\ref{Mzr}), (\ref{G2bbprawdziwe}) and (\ref{rho111}).

In this case, the resolution, $\sigma_{zd}$ has a negligible effect on the width and will therefore be ignored.
We obtain $g^{(2)}_{op}(0) -1 \simeq 0.087$.
This is in poor agreement with the experimental value, $g^{(2)}_{op}(0) -1 = 0.16$.
However the value of  $g^{(2)}_{op}(0,0) -1 $ depends crucially on the total number of particles in the quasicondensate,
and the experimental value of the total number of particles has an uncertainty of about a factor of 2. 
Therefore, we seek the value of $N$ which makes the theoretical and experimental $g^{(2)}_{op}(0,0)$ agree. 
We find $N \simeq 7 \times 10^4$, which is within the experimental uncertainty.
From this value for $N$, we can deduce
$\sigma_r/a_{hor} \simeq 1.65\ \mu\mathrm{m} $,  
$\sigma_z \simeq 0.26\ $mm, $\tilde \omega  \simeq 1.02\, \omega_r $,
$l_\phi = 92\ \mu\mathrm{m}$, $\tilde \sigma_z \simeq 1.23 \,\sigma_z$ and $\Phi_0 \simeq 0.85$.
We remind the reader that $a_{hor} = \sqrt{\hbar/m \omega_r}$ is a harmonic oscillator length.

\begin{figure}[htb]
  \centering 
  \includegraphics[width=\columnwidth]{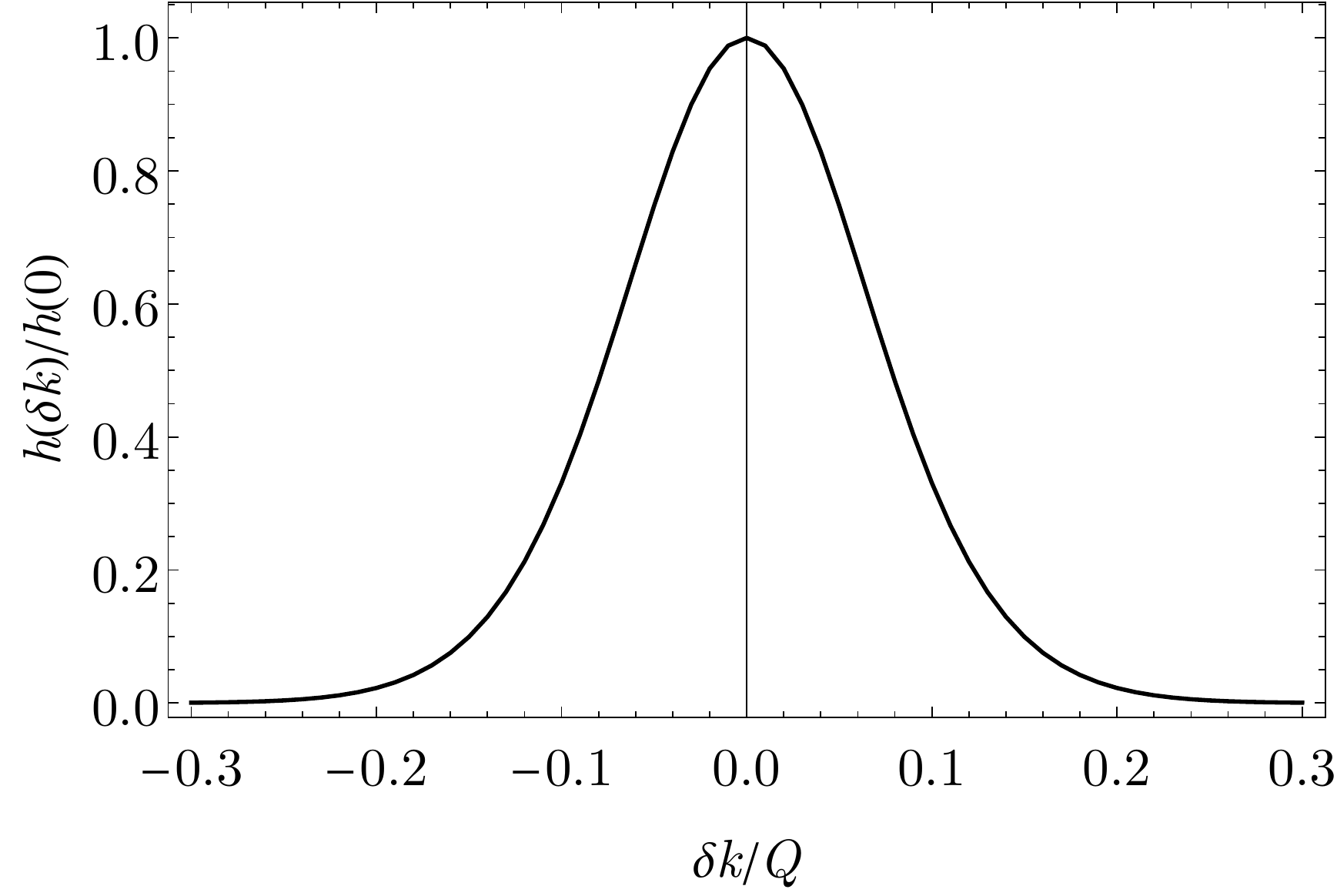}
  \caption{The radial density profile $h(\delta k)/h(0)$ of the collision sphere as a function of the radial momentum in units of $Q$.  The half width roughly equals
    $0.08\,Q$.}
\label{fig1}
\end{figure}

With the found value of $N$, we now calculate the other interesting characteristics.
We start with the radial density profile, given by Eqs.~(\ref{rho111}) and (\ref{h111}).
In Fig.~\ref{fig1}, we plot the function $h(\delta k)$ normalized to unity.
We find $h(\delta k)$ half-width of about $0.08\, Q$.
The radial profile was not measured in Ref.~\cite{paryz3},  but similar experiment, described in Ref.~\cite{karen0}, has also found a value $0.08 Q$ for this width.

\begin{figure}[htb]
  \centering
  \includegraphics[width=0.48\textwidth]{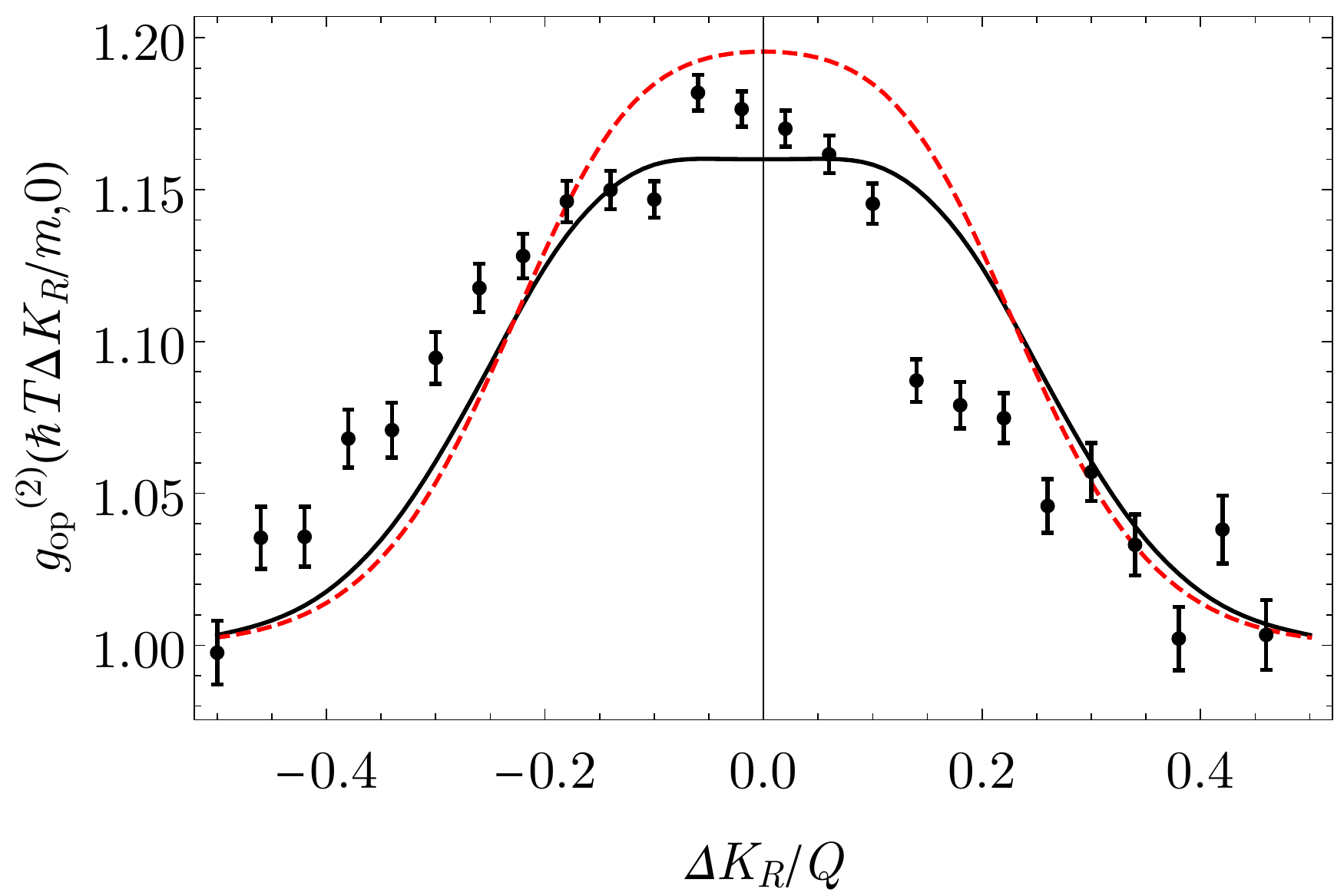}\\ \vspace{0.4cm}
  \includegraphics[width=0.48\textwidth]{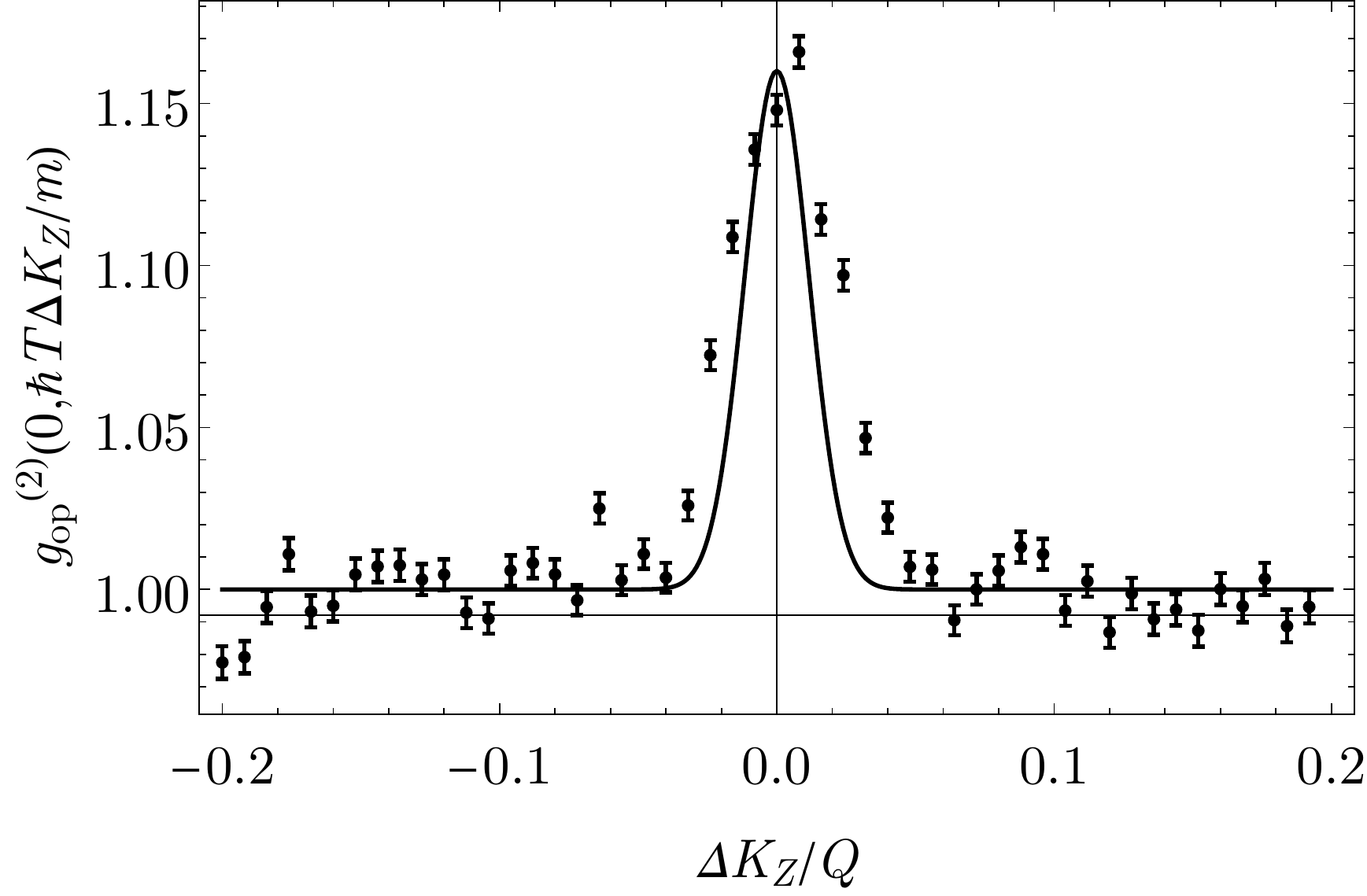}
  \caption{The function $g^{(2)}_{op}\left( \frac{\hbar T}{m} \Delta K_R, 0 \right)$ (upper panel) and $g^{(2)}_{op}\left(0, \frac{\hbar T}{m} \Delta K_Z\right)$ (lower
    panel) together with the experimental points as a function of $\Delta K_{R}/Q$ and $\Delta K_{Z}/Q$, respectively.  The red dashed line shows $g^{(2)}_{op} $ with the
    interaction term $2g|\psi_{qc}(\x,t)|^2$ present in Eq.~(\ref{H0}) neglected.}
  \label{figbb}
\end{figure}

Next, we move to $g^{(2)}_{op}(\Delta R, \Delta Z) $, the spatial dependence of the opposite correlation function.  In Fig.~\ref{figbb}, we plot $g^{(2)}_{op} \left(
\frac{\hbar T}{m} \Delta K_R , 0 \right) $ and $g^{(2)}_{op}(0, \frac{\hbar T}{m} \Delta K_Z) $ together with the experimental points.  While leaving the collision
volume, the atoms interact with the quasicondensate atoms via the mean-field potential $2g|\psi_{qc}(\x,t)|^2$ [present in the $H_0(\x,t)$ given by Eq.~(\ref{H0})].  To
see the influence of this interaction on the correlation function, we also plot $g^{(2)}_{op}$ with $2g|\psi_{qc}(\x,t)|^2$ neglected [red dashed line in the upper panel
  of Fig.~(\ref{figbb})].  The correlation in the radial direction (along $\Delta K_R$) measured in the experiment is quite close to our calculation, and we obtain
slightly better agreement by including the mean-field.  The width in the radial direction is given by the width in velocity of the colliding condensates in that
direction, that is, proportional to $1/\sigma_r$, the inverse of the radial condensate size.

In the case of the longitudinal ($\Delta K_Z$) correlations, the theoretical width is smaller than the experimental one but not much.  Unlike in the radial direction, the
longitudinal width is approximately proportional to the size of the condensate in that direction.  This is because for this correlation function, the observation does not
take place in the far field (see Appendix~\ref{ap33}).  The remaining discrepancies may be due to the use of variational approach.  The $\Delta K_Z$ correlations are
mostly given by the spatial size of the quasicondensate.  The variational ansatz decreases the spatial width in the $z$-direction compared to the true GP solution, and
therefore overestimates the $\Delta K_Z$ width.  Thus, using the GP solution the discussed difference would be smaller.

\begin{figure}[htb]
  \centering
  \includegraphics[width=0.48\textwidth]{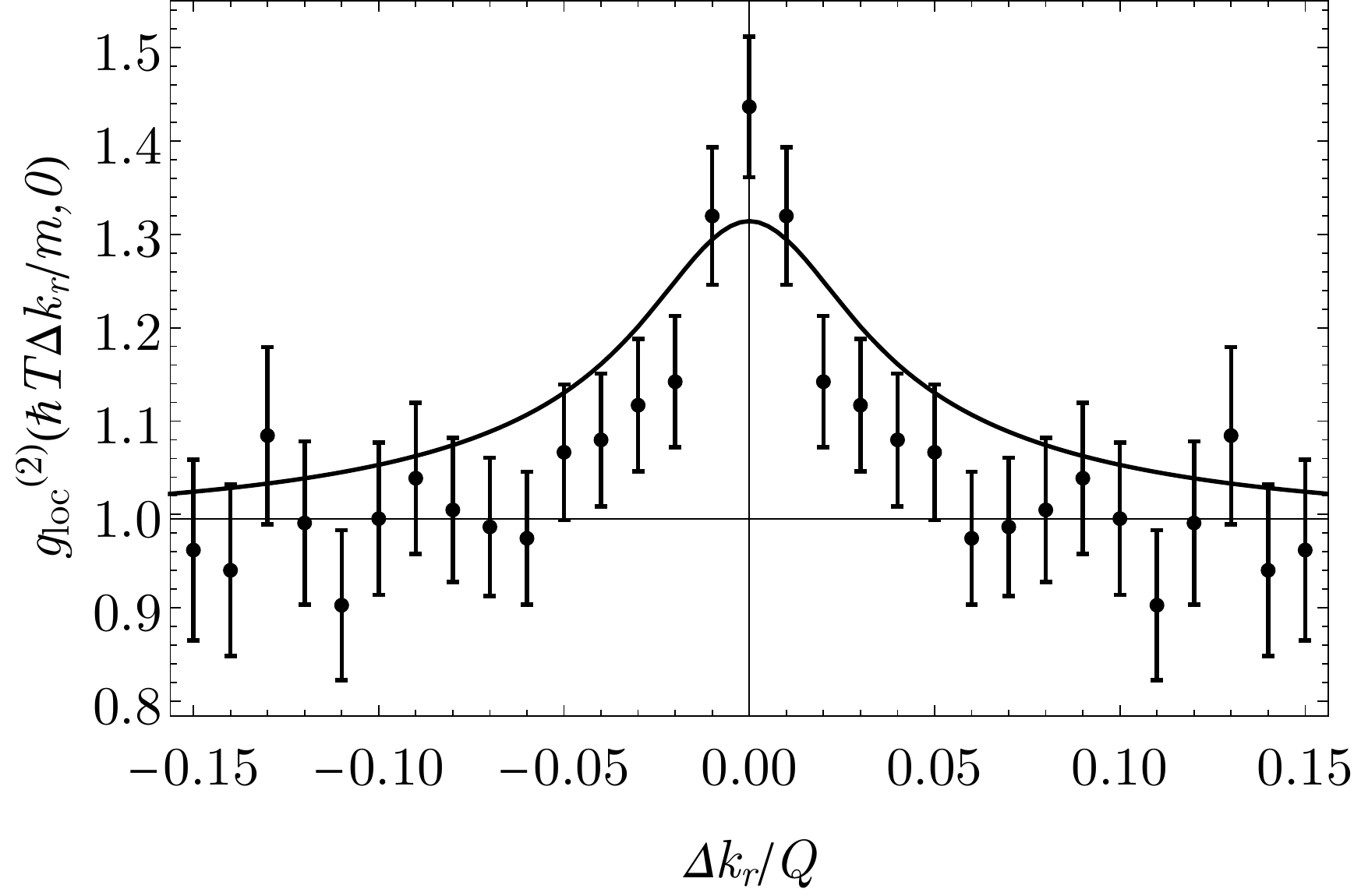}\\ \vspace{0.4cm}
  \includegraphics[width=0.48\textwidth]{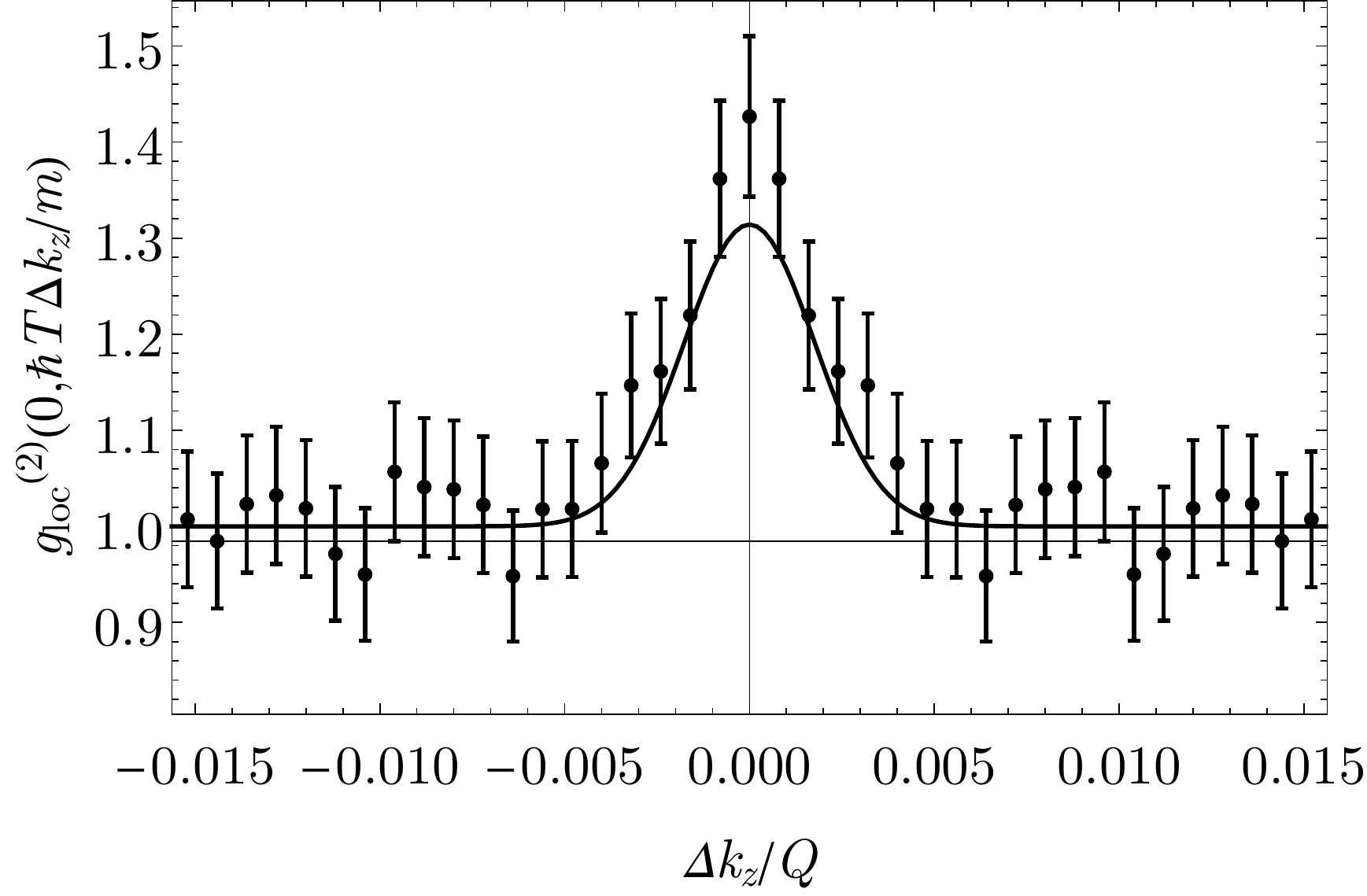}
  \caption{The function $g^{(2)}_{loc}\left( \frac{\hbar T}{m} \Delta k_r, 0 \right)$ (upper panel) and $g^{(2)}_{loc}\left(0, \frac{\hbar T}{m} \Delta k_z\right)$ (lower
    panel) together with the experimental points as a function of $\Delta k_{r}/Q$ and $\Delta k_{z}/Q$, respectively. }
  \label{figcol}
\end{figure}

Finally, we examine the local correlation $g^{(2)}_{loc}(\Delta r, \Delta z) $.  From Eqs.~(\ref{fwynik}), (\ref{G1rrr}), (\ref{rho111}), (\ref{h111}) and (\ref{w}) we
numerically calculate $g^{(2)}_{loc}$ function.  The local correlations are narrower than the opposite ones, and, therefore, the detector resolution is not negligible.
The vertical resolution $\sigma_{zd}$ is not well known, but we can find its value by fitting the data shown in Fig.~\ref{figcol}(lower panel).  We find $\sigma_{zd} =
41\ \mu \mathrm{m}$ or approximately $0.0015 v_0$ consistent with the limit set in Ref.~\cite{tunableParis}.  In Fig.~(\ref{figcol}), we plot $g^{(2)}_{loc}\left(
\frac{\hbar T}{m} \Delta k_r, 0 \right)$ and $g^{(2)}_{loc}\left(0, \frac{\hbar T}{m} \Delta k_z \right)$ together with the experimental measurements.

The theoretical correlation function is in good agreement with the experimental result in the longitudinal direction while in the radial direction it is larger by a
factor of 1.5.  This is a marked improvement over the work of Ref.~\cite{zin4}, in which we estimated much smaller radial and longitudinal widths.  This improvement is
due to the inclusion of the condensate expansion in the problem.  As the condensate expands, the density declines to the point where the collision rate becomes
negligible.  Therefore, the effective duration of the collision is shorter, and this in turn increases the energy uncertainty of the collision products.  The energy
uncertainty broadens the correlation function relative to that calculated in Ref.~\cite{zin4}, and it is still this uncertainty that determines the radial size of the
correlation, rather than the spatial width of the source.  In the case of the $z$-direction, and under the experimental conditions, the correlation width is limited by
the inverse of the size of the source, and in part also by the detector resolution.  The calculated amplitude $ g^{(2)}_{loc}(0,0)$ is somewhat smaller than the measured
one.  As in the opposite correlation case, this may be due to the use of the variational ansatz.  The fact that $ g^{(2)}_{loc} (0,0)< 2$ comes from the finite detector
resolution.  In the case of perfect detector resolution, i.e., $\sigma_{rd}=\sigma_{zd} =0$, we have $ g^{(2)}_{loc}(0,0) = 2$.


\section{Summary}\label{sec7}

We have provided an analytical treatment of the production of atom pairs during the collision of two BEC's via four-wave mixing and compared the results to an experiment.
The calculation represents a significant improvement over our previous analytical calculation~\cite{zin4}, and the agreement with the experiment is as good as an earlier
numerical treatment~\cite{karen0}.  Compared to the numerical approach, the present calculation has the advantage that we can identify the physical processes which affect
the widths and amplitudes of the correlation functions.  Compared to the earlier analytic treatment, we are now able to take into account the expansion of the condensate
during the collision.  The decrease in the density caused by the expansion is the mechanism which governs the collision time.  Therefore, the uncertainty in the energy of
the pairs is more accurately accounted for especially in the collinear correlation functions.

We are also able to clearly identify the role of the far-field condition, i.e., the condition that the time of flight be longer than ${2 m l^2 / \hbar}$, where $l$ is a
characteristic size in the collision volume.  Using the correlation length in the quasi-condensate as a characteristic size, the experiment is not in the far field for
observation is the longitudinal $z$-direction.  Thus, the correlation along $z$ for opposite pairs does not reflect the correlations in momentum, but rather more nearly
the spatial correlations.  For collinear pairs, the local correlation effect is simply a variant of the Hanbury-Brown-Twiss correlation, and the far field condition plays
no role.  For example, earth is not in the optical far field of typical stars in our galaxy.  We thus find, as expected, that apart from broadening by the detector
resolution, the local, longitudinal correlation width is a measure of the size of the source. The local correlation result also allows us to infer the detector
resolution, which is in agreement with other upper limits we have already set.

We also took into account the interaction of the scattered atoms with the colliding clouds' mean-field potential.  However, this did not affect the results much.

Finally, our treatment has included the fact that, in the experiments, the colliding clouds were quasicondensates, having a correlation length in the longitudinal
direction smaller than the condensate itself.  This feature, however, has not proved crucial for understanding the observations.

\begin{acknowledgments}
Pawe\l{} Zin acknowleges the support of Mobility Plus program.
D. Boiron and C. I. Westbrook acknowledge funding by the QuantERA grant 18-QUAN-0012-01 (CEBBEC) and the ANR grant 15-CE30-0017 (HARALAB).
\end{acknowledgments}


\appendix

\section{Temporal phase diffusion of the quasicondensate}\label{Dodatekfaza}

It follows directly from Ref.~\cite{gora} that 
\begin{eqnarray*}
  \langle (\phi(z=0,t)-\phi(z=0,0))^2 \rangle 
\end{eqnarray*}
caused by the thermal fluctuations read
\begin{eqnarray*}
  && \langle (\phi(z=0,t)-\phi(z=0,0))^2 \rangle 
  \\
  && = \frac{1}{n_{1d} \pi} \int \mbox{d} k \, n_k \frac{\varepsilon_k}{E_k}
  \sin^2 
  \left( \frac{ \varepsilon_k t}{ \hbar} \right) 
\end{eqnarray*}
where $E_k = \hbar^2 k^2 /2m$, $\varepsilon_k = \sqrt{E_k(E_k + 2 g_{1d} n_{1d})}$
and $n_k = \left(  \exp \left(  \varepsilon_k / k_BT \right) -1   \right)^{-1}$
Calculating the above for the parameters of the system considered in the paper and taking
the expansion time $\tau_{ex} = 1/\tilde \omega$ we obtain 
$ \sqrt{ \langle (\phi(z=0,\tau_{ex})-\phi(z=0,0))^2 \rangle } \simeq 0.14 $.

\section{Derivation of the anomalous density formula}\label{app2}

In this Appendix we show that the approximate treatment introduced in \cite{zinnowy}
in the case of collision of condensates applies also for our system
The crucial step of the analysis presented in \cite{zinnowy} lies in the
approximate solution to the single particle scattering problem presented in Appendix C3 of \cite{zinnowy}.
There we deal with equation
\begin{eqnarray*}
  \left( - \frac{\hbar^2}{2m} \triangle + V_{en}(\x,t) - \frac{\hbar^2 k^2}{2m} \right) \varphi(\x,t) =0
\end{eqnarray*}
with the boundary condition given by plane wave $e^{i\K\x}$. In the above $V_{en}$ is given by Eq.~(\ref{Ven}).
Following \cite{zinnowy} we introduce $\phi$ through relation $\varphi = e^{i\K\x + i \phi(\x)}$.
Substituting this form into the above equation and expanding $\phi$ in series $\phi = \phi^{(0)} + \phi^{(1)} + \ldots$
we obtain
\begin{eqnarray*}
  2 \K \nabla \phi^{(0)} &=& -\frac{2m}{\hbar^2} V_{en}
  \\
  2\K \nabla \phi^{(1)} &=& - \left( \nabla \phi^{(0)}  \right)^2 + i \triangle \phi^{(0)}
\end{eqnarray*}
with formal solution
\begin{equation} \label{phisol}
  \phi^{(j)}(\x,t) = - \frac{1}{2k} \int_{-\infty}^0 \mbox{d}s \, W_j \left(\x+s {\bf e}_\K,t \right)
\end{equation}
where $W_0 = \frac{2m}{\hbar^2} V_{en}$ and $W_1 = \left( \nabla \phi^{(0)} \right)^2 - i \triangle \phi^{(0)}$. We clearly see that $\phi^{(0)}$ depends on the $V_{en}$
potential given by Eq.~(\ref{Ven}).  We substitute to this equation $\psi_c$ given by variational ansatz and presented in Eq.~(\ref{psicN}).  As a result in the case
${\bf e}_\K = {\bf e}_x$, $t=0$ and $z=0$ we obtain
\begin{eqnarray*}
  \phi^{(0)}(x,y,0) &=& -\frac{2mgn_0 \sigma_r}{\hbar^2 Q}  \exp\left(  - \frac{y^2}{\sigma_r^2} \right) 
  \\
  &\times&
  \frac{\sqrt{\pi}}{2}
  \left( 1 + \mbox{Erf} \left(\frac{x}{\sigma_r} \right) \right)  
\end{eqnarray*}
where $n_0= N/(\pi^{3/2} \sigma_r^2 \sigma_z)$.  Inserting experimental values we obtain $\phi^{(0)}(\infty,0,0) \simeq 1.5$.  We insert the above solution into
Eq.~(\ref{phisol}) and obtain the analytic form of $\phi^{(1)}(x,y,0)$.  We find that $\phi^{(1)}(x,y,0)$ increases to infinity with the increase of $x$.  This is well
known phenomena in semiclassical approximation caused by the presence of caustics.  Anyway as stated in \cite{zinnowy} we need the approximate solution only in the space
where the cloud is present.  This gives us roughly $x<2\sigma_r$. In this part of space we find the maximal value of $\phi^{(1)}$ equal to $0.15$. As this value is
smaller than unity and also much smaller than $\phi^{(0)}$ we neglect it.  Thus we have
\begin{eqnarray*}
  \varphi_\K \simeq \exp \left( i\K\x + i \phi^{(0)}(\x) \right).
\end{eqnarray*}
In \cite{zinnowy} using the above formula we derived expression for the anomalous density given by Eqs.~(\ref{Mk2}) and (\ref{Mrealspace}).

\section{Far field conditions}\label{ap33}

In this Appendix we discuss far field conditions for $G^{(2)}_{op}$ 
function.
By far field we mean, that the correlation function  in position space 
is given by the one in momentum space, i.e.
\begin{eqnarray*}
  G_{op}^{(2)}(\x_1,\x_2,T) \propto G_{op}^{(2)}(\K_1,\K_2,T)
\end{eqnarray*}
with $\K_{1,2} = \frac{m}{\hbar T} \x_{1,2}$.
This always happens for the freely expanding gas (which is the case analyzed here) for sufficiently
long time $T$.
Here we want to find how large $T$ has to be.
First we analyze $G^{(2)}_{op} = |M|^2$ function for the condensate case.
We have
\begin{eqnarray*}
  && M(\x_1,\x_2,T) = \frac{1}{(2\pi)^3} \int \mbox{d}\K_1 \mbox{d}\K_2 \, \exp \left( i \K_1\x_1 + i \K_2 \x_2 \right)
  \\
  && \exp \left(  - i \frac{\hbar^2}{2m}(k_1^2+k_2^2) T  \right) M(\K_1,\K_2,T)
\end{eqnarray*}
It is convenient to change the variables $\X = \frac{\x_1-\x_2}{2}$, $\DX = \x_1+\x_2$,
$\KK = \frac{\K_1-\K_2}{2}$, $\DK = \K_1+\K_2$. Notice that $|\DX|\ll |\X|$ and $|\DK| \ll |\KK|$
as we deal here with opposite momentum correlations.
Than the above takes the form
\begin{eqnarray} \label{Mwzor}
  && M(\X,\DX,T) =  \frac{1}{(2\pi)^3}
  \int \mbox{d} \KK \mbox{d} \DK \, \exp \left( i 2 \KK \X   \right)
  \\ \nonumber
  && \exp \left( i \frac{\DK \DX}{2}- i\frac{\hbar }{m} \left( K^2 + \frac{\Delta K^2}{4} \right) T  \right)
  M(K,\DK,T)
\end{eqnarray}
Following \cite{zinnowy} we take the simple but realistic model of anomalous density
\begin{eqnarray} \nonumber
  M(K,\DK,T) &=& M_0(T) \exp \left(- \frac{(K-Q)^2}{2 \sigma_K^2} \right)
  \\ \label{Mk}
  & &
  \exp \left( - \frac{\Delta K_r^2}{2 \sigma_{Kr}^2} + \frac{ \Delta K_z^2}{ 2\sigma_{Kz}^2 } \right)
\end{eqnarray}
where we take $\sigma_{Kr} \approx \frac{\sqrt{2}}{\sigma_r}$, $\sigma_{Kz} \approx \frac{\sqrt{2}}{\sigma_z}$,
$\sigma_K  \approx (Qa_{hor}^2)^{-1} $. 
Inserting  Eq.~(\ref{Mk}) into Eq.~(\ref{Mwzor}) we arrive at
\begin{eqnarray} \label{m11}
  && M(\X,\DX,T) =  M_0 M_1(\X,T) M_2(\DX,T)
  \\ \nonumber
  \\ \nonumber
  && M_1(\X,T) = \int \frac{\mbox{d} \KK}{(2\pi)^{3/2}} \, \exp \left( i 2 \KK \X - i\frac{\hbar  K^2}{m} T \right)
  \\ \nonumber
  && \times \exp \left(  - \frac{(K-Q)^2}{2 \sigma_K^2}   \right)
  \\ \nonumber
  \\ \nonumber
  && M_2(\DX,T) =  \frac{1}{(2\pi)^{3/2}} \int \mbox{d} \DK \, \exp \left( i \frac{\DK \DX}{2}\right)
  \\ \nonumber
  && \times \exp \left( - i\frac{\hbar  \Delta K^2}{4 m}  T- \frac{\Delta K_r^2}{2 \sigma_{Kr}^2} + \frac{ \Delta K_z^2}{ 2\sigma_{Kz}^2 } \right)
\end{eqnarray}
As the above integrals are of gaussian form they can be integrated analytically to get
\begin{eqnarray*}
  M_1(\X,T) &=& \frac{Q \sigma_K}{2iR} \frac{1}{\sqrt{ 1 + i\frac{2\hbar \sigma_K^2 T}{m} }} \exp \left(- i\frac{\hbar  Q^2}{m} T  \right)
  \\
  & &
  \times \exp \left( 2i Q R  - \frac{ 2(R - v_0T)^2 \sigma_K^2}{1 + i\frac{2\hbar \sigma_K^2 T}{m} }  \right)
\end{eqnarray*}
and
\begin{eqnarray*}
  && M_2(\DX,T) =  \left(1 + i \frac{\hbar \sigma_{Kr}^2 T}{2m} \right)^{-1}
  \left( 1 + i \frac{\hbar \sigma_{Kz}^2 T}{2m} \right)^{-1/2}
  \\
  &&
  \times \exp \left(  -\frac{(\Delta X^2+\Delta Y^2) \sigma_{Kr}^2}{8 \left(1 + i \frac{\hbar \sigma_{Kr}^2 T}{2m} \right)}
  -\frac{\Delta Z^2\sigma_{Kz}^2}{8 \left(1 + i \frac{\hbar \sigma_{Kz}^2 T}{2m} \right)}
  \right).
\end{eqnarray*}
In Eq.~(\ref{Mwzor}) we notice two independent propagators for $\DX$ and $\X$.  The condition to be in the far field for $\Delta K_{x,y}$ is $\frac{\hbar \sigma_{Kr}^2
  T}{2m} \gg 1$.  For $\Delta K_z$ it is $\frac{\hbar \sigma_{Kz}^2 T}{2m} \gg 1$ and for $K$ is $\frac{2\hbar \sigma_K^2 T}{m} \gg 1$.

Analyzing the above derivation we may additionally read the condition to be in the ``near field,'' i.e.,  when $\frac{\hbar \sigma_{Kr,Kz,K}^2 T}{2m} \ll 1$. Then the
correlations are $T$ independent.  We note that in the case $\sigma_{Kr} \gg \sigma_{Kz}$ the $\Delta K_{x,y}$ correlation may reach the far field, where as $\Delta K_z$
may be still in the near field.  If we consider the condensate regime of our system and substitute the approximate values (see \cite{zinnowy}) $\sigma_{Kr} \approx
\frac{\sqrt{2}}{\sigma_r}$, $\sigma_{Kz} \approx \frac{\sqrt{2}}{\sigma_z}$, $\sigma_K \approx \frac{1}{Q a_{hor}^2} $, $T=0.3$s we find
\begin{eqnarray*}
  \frac{2\hbar \sigma_K^2 T}{m} \simeq 100 \ \ \  \frac{\hbar \sigma_{Kr}^2 T}{2m} \simeq 1.6 \times 10^3 \ \ \  
  \frac{\hbar \sigma_{Kz}^2 T}{2m} \simeq 0.06
\end{eqnarray*}
These results give the above described case.

The above was derived for the condensate collision case.  Still we can use it in the quasicondensate case.  The $\sigma_{Kr}$ and $\sigma_{Kz}$ present in Eq.~(\ref{Mk})
are directly connected to the momentum widths of the initial wave packets \cite{zinnowy}.  In the condensate case it was $\sigma_{Kr} \approx \frac{\sqrt{2}}{\sigma_r}$,
$\sigma_{Kz} \approx \frac{\sqrt{2}}{\sigma_z}$. In the quasicondensate case we can take $\sigma_{Kz} \approx 1/l_{\phi}$. As $l_\phi \gg a_{hor}$ this does not change
the value of $\sigma_K$ \cite{zinnowy}.  Therefore in the quasicondensate as in the condensate case the far field is reached for $\Delta K_{x,y}$ and $K$ variables.  In
the $\Delta K_z$ variable we obtain $ \frac{\hbar T \sigma_{Kz}^2}{2m} \simeq 0.16$ which shows that this correlation is not in the far field.

\section{The calculation of $G_{op}^{(2)}$ function.} \label{G2bbApp}

In Appendix \ref{ap33} we show that in the experiment the far field is reached 
in the case  of $\Delta X $, $\Delta Y$ and $\X$ variables.
In such a case the anomalous density given by Eq.~(\ref{Mrealspace}) takes the form
\begin{eqnarray} \nonumber
  && M(\X,\DX,T) =  \exp \left(   i \left( K^2 + \frac{\Delta K_x^2 +\Delta K_y^2}{4}  \right) \frac{\hbar  T}{m} \right) 
  \\ \nonumber
  &&
  \left( \frac{ m}{i \hbar T} \right)^{5/2}
  \frac{1}{ 2 \sqrt{\pi}}
  \int  \mbox{d} \Delta K_z
  \\ \label{m2x}
  && \exp \left( i \Delta K_z \frac{\Delta Z}{2}- i\frac{\hbar \Delta K_z^2}{4m} T  \right)
  M(\KK,\DK)
\end{eqnarray}
where $\Delta X = \frac{\hbar \Delta K_x}{m} T$ and $\Delta Y = \frac{\hbar \Delta K_y}{m} T$, $\X = \frac{\hbar \KK}{m}T$.
From Eqs.~(\ref{Mk2}), (\ref{Bmf}) and (\ref{Ba}) we obtain
\begin{eqnarray} \nonumber
  && M(\KK,\DK)
  = \frac{ g }{i\hbar (2\pi)^3}   \int_0^\infty \mbox{d} t \int \mbox{d} \x \, \psi_c^2 (\x,t)\exp \left( - i \DK \x   \right)
  \\ \label{m2k}
  &&
  \exp \left( i \frac{\hbar}{m} \left( K^2 - Q^2 + \frac{\DK^2}{4} \right)  t +2i \phi(z) -i \Phi \right).
\end{eqnarray}
As written in the main body of the paper in the experiment the measured quantity is $G^{(2)}_{op} \left( \X,\DX,T\right)$ averaged over position $\X$ i.e.
\begin{eqnarray}\nonumber
  G^{(2)}_{op} \left(\DX,T\right) &=& \int \mbox{d} \X \, G^{(2)}_{op} \left( \X,\DX,T\right)
  \\ \nonumber
  &=& \int \mbox{d} \X \, \langle |M(\X,\DX,T)|^2 \rangle_{cl}.
\end{eqnarray}
Substituting  Eqs.~(\ref{m2x}) and (\ref{m2k}) into the above we obtain
\begin{widetext}
  \begin{eqnarray} \nonumber
    && G^{(2)}_{op} \left(\DX,T\right)  \simeq \left( \frac{m}{\hbar T} \right)^2 \frac{m  Q}{\hbar^3}
    \frac{1}{4(2\pi)^6}  \int_0^\infty \mbox{d} t \int \mbox{d} \Omega_\KK
    \\ \label{G2bbG2}
    &&
    \langle \left|
    \int  \mbox{d} \Delta K_z \,
    \exp \left( i \Delta K_z \frac{\Delta Z}{2}- i\frac{\hbar \Delta K_z^2}{4m} T  \right)
    \int \mbox{d} \x \,  \exp \left(  - i \DK \x \right) g\psi_c^2(\x,t) \exp \left(2i\phi(z)-i\Phi \right)
    \right|^2 \rangle_{cl}
  \end{eqnarray}
\end{widetext}
where  $ \mbox{d} \X = \left(\frac{\hbar T}{m} \right)^3 \mbox{d} \KK$
and we used the Dirac delta approximation described in \cite{zinnowy}.

To continue with further calculation of the $G^{(2)}_{op}$ function we now analyze the phase $\Phi$.  Due to the axial symmetry of the system we may take ${\bf e}_\KK =
(\sin \theta,0,\cos \theta)$.  As $\sigma_r \ll \sigma_z $ and $\sin \theta > \frac{\sqrt{3}}{2}$ we approximate
\begin{eqnarray*}
  \Phi(\x,{\bf e}_{\KK},t) &\simeq& \frac{1}{\sin \theta} \Phi(\x,{\bf e}_x,t)
  \\
  & &
  = \frac{1}{\sin \theta} \frac{m}{\hbar^2 Q} \int^\infty_{-\infty} \mbox{d} s \, 2g|\psi_c(x+s,y,z,t)|^2
\end{eqnarray*}
where we used Eqs.~(\ref{MFPhi}) and (\ref{Ven}).
Inserting into the above the variational ansatz solution  of $\psi_c$ given by Eq.~(\ref{psicN})
we obtain
\begin{equation}\label{Phiup}
  \Phi(y,z,\theta) \simeq 
  \frac{1}{\sin \theta} \frac{2g m}{\hbar^2 Q} \frac{n_0 \sigma_r^2}{\sigma_r(t)} \sqrt{\pi} \exp\left( - \frac{y^2}{\sigma_r^2(t)} - \frac{z^2}{\sigma_z^2}\right)
\end{equation}
We find that the gradient of the phase $\phi(z)$ is much larger than $\partial_z \Phi$.
This makes us to approximate the integral present in Eq.~(\ref{G2bbG2}) as
\begin{eqnarray} \nonumber
  && \int \mbox{d} z \,  \exp \left(  - i \Delta K_z z \right)  \exp \left( - \frac{z^2}{\sigma_z^2} +2i\phi(z)
  -i\Phi(y,z,\theta,t) \right)
  \\ \label{row123}
  && \approx \exp \left( - i \widetilde \Phi(y,\theta,t) \right) 
  \int \mbox{d} z \,  \exp \left(  - i \Delta K_z z  - \frac{z^2}{\sigma_z^2} +2i\phi(z) \right).
\end{eqnarray}
In the above we  used the variational ansatz wave-function $\psi_c$ given by Eq.~(\ref{psicN}).
We also introduced $\widetilde \Phi(y,\theta,t)$ which is $\Phi(y,z,\theta,t)$ averaged over $z$ with the 
condensate density $ \exp \left( - \frac{z^2}{\sigma_z^2} \right) $:
\begin{eqnarray*}
  \widetilde \Phi(y,\theta,t) = 
  \frac{\int \mbox{d} z \, \Phi(y,z,\theta,t) \exp \left( - \frac{z^2}{\sigma_z^2} \right)}{ \int \mbox{d} z \, \exp \left( - \frac{z^2}{\sigma_z^2} \right) }.
\end{eqnarray*}
Substituting $\Phi$ given by Eq.~(\ref{Phiup}) we obtain:
\begin{equation}\label{Phiav}
  \widetilde \Phi(\theta,y,t) \simeq \Phi_0 \frac{1}{\sin \theta \sqrt{1+\tilde \omega^2 t^2}} \exp\left( - \frac{y^2}{\sigma_r^2(t)}\right)
\end{equation}
where $\Phi_0 =  \frac{2g m n_0 \sigma_r}{\hbar^2 Q} \sqrt{\frac{\pi}{2}} \simeq 1.07 $.

Inserting the approximation given by Eq.~(\ref{row123}) into Eq.~(\ref{G2bbG2}) we notice 
the presence of the function defined as
\begin{eqnarray*}
  M_z(\Delta Z,T) &=& \int  \mbox{d} \Delta K_z \,
  \exp \left( i \Delta K_z \frac{\Delta Z}{2}- i\frac{\hbar \Delta K_z^2}{4m} T   \right)
  \\
  & & \int \mbox{d}z \exp \left( - i \Delta K_z z  + 2 i \phi(z) -  \frac{z^2}{\sigma_z^2}\right).
\end{eqnarray*}
We notice that the above can be rewritten as
\begin{equation}\label{Mz11}
  \langle |M_z(\Delta Z,T)|^2 \rangle_{cl} = \int   \mbox{d} K_z \, W \left( \Delta Z - \frac{\hbar T}{m} K_z, K_z  \right)
\end{equation}
where
\begin{eqnarray*}
  && W(\Delta Z, K_z) = 2\pi \int  \mbox{d} \Delta z \, 
  \exp \left( - i K_z \Delta z -\frac{\Delta Z^2 +\Delta z^2}{2\sigma_z^2} \right)
  \\
  \\
  && \times 
  \langle \exp \left( 2 i \phi \left(\frac{\Delta Z + \Delta z}{2} \right) -2 i \phi \left(\frac{\Delta Z - \Delta z}{2} \right)  \right)
\rangle_{cl}
\end{eqnarray*}
is a Wigner function.  Using the local density approximation we have
\begin{equation}\label{W11}
  W(\Delta Z, K_z) \simeq   \exp \left( -\frac{\Delta Z^2 }{2\sigma_z^2} \right) \rho(K_z)
\end{equation}
where
\begin{eqnarray*}
  && \rho(K_z) = 2\pi \int  \mbox{d} \Delta z \, 
  \exp \left( - i K_z \Delta z \right)
  \\
  \\
  && \times
  \langle \exp \left( 2 i \phi \left(\frac{\Delta z}{2} \right) -2 i \phi \left(-\frac{\Delta z}{2} \right)  \right)
  \rangle_{cl}.
\end{eqnarray*}
The above can be calculated using the formulas for 1D uniform system  \cite{dima1}
arriving at
\begin{eqnarray*}
  \rho(K_z) = \frac{2\pi l_\phi}{1+(K_z l_\phi /2)^2} 
\end{eqnarray*}
where $l_\phi$ is the thermal coherence length given by Eq.~(\ref{d}).
Introducing the function $G_z^{(2)}(\Delta Z,T)$  Eq.~(\ref{Mz11}) can be rewritten as
\begin{eqnarray*}
  G_z^{(2)}(\Delta Z,T) =  \int   \mbox{d} K_z \, G_z^{(2)} \left( \Delta Z - \frac{\hbar T}{m} K_z, 0  \right) \rho(K_z)
\end{eqnarray*}
where we used Eq.~(\ref{W11}) and $ G_z^{(2)} \left( \Delta Z, 0  \right) =   \exp \left( -\frac{\Delta Z^2 }{2\sigma_z^2} \right) $.

From Eqs.~(\ref{G2bbG}), (\ref{row123}), (\ref{psicN}), (\ref{G2bbG2}) and (\ref{Mzr})
we obtain 
\begin{widetext}
  \begin{eqnarray} \label{G2bbprawdziwe}
    G_\rho^{(2)}(\Delta R,{\bf e}_\KK,t) &=& C \frac{\pi}{ (1+\tilde t^2) \sqrt{ 1 + \tilde t^2 \frac{\sigma_r^4}{\tilde a_{hor}^4} } }
    \exp \left( - \frac{1}{2} \Delta K^2 \cos^2 \phi \frac{1+\tilde t^2}{1 + \tilde t^2 \frac{\sigma_r^4}{\tilde a_{hor}^4}}
    \right) 
    \\ \nonumber	
    & &	\left|
    \int  \mbox{d} \tilde y \,  \exp \left(  - i \Delta K \sin \phi \tilde y -i\Phi(\theta,\sigma_r \tilde y,\tilde t/\tilde \omega)
    -\frac{\tilde y^2}{1+\tilde  t^2 } \left( 1 - i \tilde  t \frac{\sigma_r^2}{\tilde a_{hor}^2} \right)  \right)
    \right|^2
  \end{eqnarray}
\end{widetext}
where $\Delta K = \frac{m \sigma_r \Delta R}{\hbar T} $, $\tilde t = \tilde \omega t$,
$\tilde y = y/\sigma_r$
and
\begin{eqnarray*}
  C &=& \left( \frac{m}{\hbar T} \right)^2 \frac{m Q}{\hbar^3} 
  \frac{1}{4(2\pi)^6} \left(g n_0 \right)^2 \sigma_r^4.
  \\
  &=&  \left( \frac{m}{\hbar T} \right)^2 \frac{1}{16 \pi^4} \frac{\hbar Q}{m} n_0^2 a^2 \sigma_r^4
\end{eqnarray*}
In the above we used standard  parametrization ${\bf e}_\KK = (\sin \theta \cos \phi, \sin\theta \sin \phi, \cos \theta)$.
It is important to mention that in the experiment the averaging over the solid angles $\Omega_\KK$
is performed only over part of the sphere i.e. for $ \frac{\pi}{3} < \theta < \frac{2\pi}{3} $ \cite{paryz3}.  

We also calculate the above neglecting the phase $\Phi$. Than it can be integrated arriving at
\begin{equation}\label{G2bbuproszczone}
  G_\rho^{(2)}(\Delta R,{\bf e}_\KK,t) =  \frac{C \pi^2}{ \left( 1 + \tilde t^2 \frac{\sigma_r^4}{\tilde a_{hor}^4} \right) }
  \exp \left( - \frac{1}{2} \Delta K^2  \frac{1+\tilde t^2}{1 + \tilde t^2 \frac{\sigma_r^4}{\tilde a_{hor}^4}}
  \right) 
\end{equation}


\begin{thebibliography}{99}

\bibitem{bellAspect} A. Aspect, Nature {\bf 398}, 189 (1999).
\bibitem{oum} C. K. Hong, Z. Y. Ou, and L. Mandel, Phys. Rev. Lett. {\bf 59}, 2044 (1987).


\bibitem{ghostphotons} Baris I. Erkmen and Jeffrey H. Shapiro,  Adv. Opt. Photon. {\bf 2}, 405 (2010).



\bibitem{tunableParis} M. Bonneau, J. Ruaudel, R. Lopes, J.-C. Jaskula, A. Aspect, D. Boiron, and C. I. Westbrook, Phys. Rev. A {\bf 87}, 061603(R) (2013).

\bibitem{viennaTB} R. B\"ucker, J. Grond,S. Manz, T. Berrada, T. Betz, C. Koller, U. Hohenester, T. Schumm, A. Perrin, and J. Schmiedmayer, Nat. Phys. {\bf 7}, 608 (2011).


\bibitem{keterle1} J. M. Vogels, K. Xu, and W. Ketterle, Phys. Rev. Lett. {\bf 89}, 020401 (2002).

\bibitem{paryz0} A. Perrin, H. Chang, V. Krachmalnicoff, M. Schellekens, D. Boiron, A. Aspect, and C. I. Westbrook, Phys. Rev. Lett. {\bf 99}, 150405 (2007).

\bibitem{paryz1} V. Krachmalnicoff, J.-C. Jaskula, M. Bonneau, V. Leung, G. B. Partridge, D. Boiron, C. I. Westbrook, P. Deuar, P. Zi\'n, M. Trippenbach, and K. V. Kheruntsyan, 
Phys. Rev. Lett. {\bf 104}, 150402 (2010).


\bibitem{paryz2} J.-C. Jaskula, M. Bonneau, G. B. Partridge, V. Krachmalnicoff, P. Deuar, K. V. Kheruntsyan, A. Aspect, D. Boiron, and C. I. Westbrook, 
Phys. Rev. Lett. {\bf 105}, 190402 (2010).


\bibitem{wasakdeuar} P. Deuar, T. Wasak, P. Zi\'n, J. Chwede\'nczuk, and M. Trippenbach, Phys. Rev. A {\bf 88}, 013617 (2013).


\bibitem{paryz3}
K. V. Kheruntsyan, J.-C. Jaskula, P. Deuar, M. Bonneau, G. B. Partridge, J. Ruaudel, R. Lopes, D. Boiron, and C. I. Westbrook, Phys. Rev. Lett. {\bf 108}, 260401 (2012).

\bibitem{wasakCSI} T. Wasak, P. Sza\'nkowski, P. Zi\'n, M. Trippenbach, and J. Chwede\'nczuk, Phys. Rev. A {\bf 90}, 033616 (2014).

\bibitem{wasakCSI2} T. Wasak, P. Sza\'nkowski, M. Trippenbach, and J. Chwedeńczuk, Quantum Information Processing {\bf 15}, 269 (2015).
 
\bibitem{wasakEnt} Tomasz Wasak, Augusto Smerzi, and Jan Chwede\'nczuk, Scientific Reports {\bf 8}, 1777 (2018).
 

\bibitem{atomicHOMtheo} R. J. Lewis-Swan, and K. V. Kheruntsyan, Nat. Commun. {\bf 5}, 3752 (2014).

\bibitem{atomicHOM} R. Lopes, A. Imanaliev, A. Aspect, M. Cheneau, D. Boiron, and  C. I. Westbrook, Nature {\bf 520}, 66 (2015).

\bibitem{ghost}   R. I. Khakimov, B. M. Henson, D. K. Shin, S. S. Hodgman, R. G. Dall, K. G. H. Baldwin, and A. G. Truscott, Nature {\bf 540}, 100 (2016).

\bibitem{ghost2} S. S. Hodgman, W. Bu, S. B. Mann, R. I. Khakimov, and A. G. Truscott, Phys. Rev. Lett. {\bf 122}, 233601 (2019).

\bibitem{wasakBellCorr} D. K. Shin, B. M. Henson, S. S. Hodgman, T. Wasak, J. Chwedenczuk, A. G. Truscott, arXiv:1811.05681 (2018).

\bibitem{gradioTruscott} D. K. Shin, J. A. Ross, B. M. Henson, S. S. Hodgman, A. G. Truscott, arXiv:1906.08958 (2019).

\bibitem{atomBell} R. J. Lewis-Swan, K. V. Kheruntsyan, Phys. Rev. A {\bf 91}, 052114 (2015).

\bibitem{wasakBell} T. Wasak, J. Chwede\'nczuk, Phys. Rev. Lett. {\bf 120}, 140406 (2018).

\bibitem{njpWasak} T. Wasak, P. Sza\'nkowski, R. B\"ucker, J. Chwede\'nczuk, and M. Trippenbach, New J. Phys., {\bf 16}, 013041 (2014).

\bibitem{hawkingZin} D. Boiron, A. Fabbri, P.-\'E Larr\'e, N. Pavloff, C. I. Westbrook, and P. Zi\'n, Phys. Rev. Lett. {\bf 115}, 025301 (2015).



\bibitem{Wieden} M. Keller, M. Kotyrba, F. Leupold, M. Singh, M. Ebner, and A. Zeilinger, Phys. Rev. A {\bf 90}, 063607 (2014).



\bibitem{jur} V. A. Yurovsky, Phys. Rev. A {\bf 65}, 033605 (2002).

\bibitem{bach} R. Bach, M. Trippenbach, K. Rz\c{a}\.zewski, Phys. Rev. A {\bf 65}, 063605 (2002).

\bibitem{zin1} P. Zi\'n,  J. Chwede\'nczuk,  A. Veitia,  K. Rz\c{a}\.zewski, M. Trippenbach, Phys. Rev. Lett. {\bf 94}, 200401 (2005).

\bibitem{zin2} P.  Zi\'n,  J.  Chwede\'nczuk,  M.  Trippenbach,  Phys. Rev. A {\bf 73}, 033602 (2006).

\bibitem{zin3}  J. Chwede\'nczuk, P. Zi\'n, K. Rz\c{a}\.zewski, M. Trippenbach, Phys. Rev. Lett. {\bf 97}, 170404 (2006).

\bibitem{deuar1} P. Deuar, and P. D. Drummond, Phys. Rev. Lett. {\bf 98}, 120402 (2007).


\bibitem{zin4} J. Chwede\'nczuk, P. Zi\'n, M. Trippenbach, A. Perrin, V. Leung, D. Boiron, C.I. Westbrook, Phys. Rev. A {\bf 78}, 053605 (2008).


\bibitem{deuar2} P. Deuar, J. Chwede\'nczuk, M. Trippenbach, and P. Zi\'n, Phys. Rev. A {\bf 83}, 063625 (2011).


\bibitem{gardiner1} A. A. Norrie, R. J. Ballagh, and C. W. Gardiner, Phys. Rev. Lett. {\bf 94}, 040401 (2005).

\bibitem{gardiner2} A. A. Norrie, R. J. Ballagh, and C. W. Gardiner, Phys. Rev. A {\bf 73}, 043617 (2006).

\bibitem{paryz10} K. Molmer, A. Perrin, V. Krachmalnicoff, V. Leung, D. Boiron, A. Aspect, C. I. Westbrook, Phys. Rev. A {\bf 77}, 033601 (2008).


\bibitem{karen0} A. Perrin, C. M. Savage, D. Boiron, V. Krachmalnicoff, C. I. Westbrook, and K. V. Kheruntsyan, New J. Phys. {\bf 10}, 045021 (2008).

\bibitem{karen1} M. Ogren, K. V. Kheruntsyan, Phys. Rev. A {\bf 79}, 021606(R) (2009).


\bibitem{zinnowy} Pawe{\l} Zi\'n, Tomasz Wasak, Phys. Rev. A {\bf 97}, 043620 (2018).

\bibitem{gora} D.S. Petrov, D.M. Gangardt and G.V. Shlyapnikov, J. Phys. IV France {\bf 116}, 5-44  (2004)
\bibitem{gerbier} F. Gerbier, Eur. Phys. Lett. {\bf 66}, 771 (2004)

\bibitem{wasakRaman} T. Wasak, J. Chwede\'nczuk, P. Zi\'n, and M. Trippenbach, Phys. Rev. A {\bf 86}, 043621 (2012).

\bibitem{sl} S. Moal, M. Portier, J. Kim, J. Dugue, U. D. Rapol, M. Leduc, and C. Cohen-Tannoudji, Phys. Rev. Lett. {\bf 96}, 023203 (2006).

\bibitem{marek} Marek Trippenbach, Y. B. Band, and P. S. Julienne, Phys. Rev. A {\bf 62}, 023608 (2000).











\bibitem{dopiska} We neglect the quantum depletion of the condensates since we are interested in the scattered particles.



%

\bibitem{HBT0}  M. Schellekens, R. Hoppeler, A. Perrin, J. Viana Gomes, D. Boiron, A. Aspect, C. I. Westbrook, Science {\bf 310}, 648 (2005).













\bibitem{dima1} I. Bouchoule, M. Arzamasovs, K. V. Kheruntsyan, and D. M. Gangardt, Phys. Rev. A {\bf 86}, 033626 (2012)



\end{thebibliography}
\end{document}